\documentclass[12pt]{article}
\usepackage{amsthm,amsmath,amssymb,bbm,bm}

\usepackage{multirow}
\usepackage[pdftex]{graphicx}
\usepackage{subfigure}
\usepackage{wrapfig}
\usepackage{CJK}
\usepackage{array}
\usepackage{url}
\usepackage{booktabs}
\usepackage{algorithm}
\usepackage{algorithmic}
\usepackage{placeins}
\usepackage{mathtools}
\usepackage{romannum}
\usepackage{mathrsfs}
\usepackage{dsfont}
\usepackage{relsize}
\usepackage{rotating}
\usepackage{enumitem}
\usepackage{setspace}
\usepackage{minitoc}
\usepackage{etoc}
\usepackage{authblk}
\usepackage[numbers]{natbib}

\usepackage[usenames,dvipsnames,svgnames,table]{xcolor}

\usepackage[colorlinks=true,
            linkcolor=red,
            anchorcolor=blue,
            citecolor=blue,
            urlcolor=blue]{hyperref}

\makeatletter
\newcommand{\vast}{\bBigg@{3}}
\makeatother

\usepackage{chngcntr}
\counterwithin{equation}{section}
\definecolor{ao(english)}{rgb}{0.0, 0.5, 0.0}
	
\theoremstyle{definition}

\counterwithin{thm}{section}

\counterwithin{rmk}{section}
\counterwithin{proposition}{section}

\counterwithin{corollary}{section}

\def\##1\#{\begin{align}#1\end{align}}
\def\$#1\${\begin{align*}#1\end{align*}}

\def\beq#1\eeq{\begin{equation}#1\end{equation}}
\def\baa#1\eaa{\begin{eqnarray}#1\end{eqnarray}}
\def\bal#1\eal{\begin{align}#1\end{align}}

\DeclareMathOperator*{\argmin}{arg\,min}
\DeclareMathOperator*{\argmax}{arg\,max}

\usepackage[margin=0.85in]{geometry}
\usepackage{ulem}

\usepackage{color}

\begin{document}
\title{A Seamless Phase II/III Design with Dose Optimization for Oncology Drug Development }
\author[1]{Yuhan Li}
\author[2]{Yiding Zhang\thanks{Correspondence to: yiding.zhang@sanofi.com}}
\author[2]{Gu Mi}
\author[2]{Ji Lin}
\affil[1]{Department of Statistics, University of Illinois Urbana-Champaign}
\affil[2]{Biostatistics and Programming, Sanofi}

\date{}

%\addbibresource{seamless_design.bib}
\maketitle
\setcounter{page}{1}
\pagenumbering{arabic}

\begin{abstract}
The US FDA's Project Optimus initiative that emphasizes dose optimization prior to marketing approval represents a pivotal shift in oncology drug development. It has a ripple effect for rethinking what changes may be made to conventional pivotal trial designs to incorporate a dose optimization component. Aligned with this initiative, we propose a novel \textbf{S}eamless Phase II/III \textbf{D}esign with \textbf{D}ose \textbf{O}ptimization (SDDO framework). The proposed design starts with dose optimization in a randomized setting, leading to an interim analysis focused on optimal dose selection, trial continuation decisions, and sample size re-estimation (SSR). Based on the decision at interim analysis, patient enrollment continues for both the selected dose arm and control arm, and the significance of treatment effects will be determined at final analysis. The SDDO framework offers increased flexibility and cost-efficiency through sample size adjustment, while stringently controlling the Type I error. This proposed design also facilitates both Accelerated Approval (AA) and regular approval in a ``one-trial'' approach.  Extensive simulation studies confirm that our design reliably identifies the optimal dosage and makes preferable decisions with a reduced sample size while retaining statistical power.
\end{abstract}

%This intern project focuses on the latter situation where we propose a phase 2/3 seamless adaptive design with stage 1 of dose optimization in a randomized setting, followed by a formal interim analysis with dual objectives of optimized dose selection and sample-size re-estimation (SSR); stage 2 then kicks off with continued enrollment of the selected dose (arm) and the concurrent control arm (enrolled since stage 1) to be compared at the final analysis. In this presentation, we will talk about motivation of the proposed design, associated statistical concepts (e.g., predictive probability of success, SSR, multiplicity controls, etc.), advantages of using a Bayesian framework, results from comprehensive simulation studies, some practical considerations, and future works. 

\section{Introduction}

Oncology drug development relies heavily on clinical trials to assess the safety and efficacy of novel therapies for cancer patients. However, conventional approaches such as separated Phase II and Phase III studies may not be the optimal option given the ever-changing competitive and treatment landscapes, the ever-increasing development costs, and the high attrition rates across different stages of development. According to Wong et al\citep{wong2019estimation}, the probability of success rate in oncology is only 57.6\% from Phase I to Phase II, and 32.7\% from Phase II to Phase III, both of which are the lowest compared to other therapeutic areas. There is a justified urgency to adopt a ``quick-to-win, fast-to-fail'' mindset and utilize more efficient designs for oncology drug development to expedite patient access to potentially life-saving treatments.

%\sout{Oncology drug development relies heavily on clinical trials to assess the safety and efficacy of novel therapies for cancer participants. However, the conventional approach of conducting separate and sequential phase II and phase III trials poses significant challenges and limitations, such as extended timelines, escalating costs, and delayed patient access to potentially life-saving treatments \citep{orloff2009future}.}

%\sout{In regard to such challenges,} 

The European Medicine Agency (EMA) Reflection Paper \cite{committee2007reflection} has a pioneering effort in shaping the regulatory direction of adaptive designs. It has been released during a period of extensive discussion on topics like methodology, implementation, and regulatory validation of adaptive designs. Since then, there has been a growing interest in adopting a more integrated and adaptive approach known as ``seamless design'' in oncology clinical trials \citep{orloff2009future,jenkins2011adaptive,chen20182}. Seamless Phase II/III trials, in particular, aim to overcome the limitations of traditional designs by seamlessly transitioning from exploratory to confirmatory phases within a single trial framework. This approach offers several potential benefits, including the concurrent recruitment of participants for both Phase II and Phase III \citep{chen20182}, and the flexibility to adjust sample size based on interim analysis while allowing conventional test procedure to be performed at the final analysis \citep{li2022flexible}. These desired features have the potential to expedite the development process and save costs while maintaining statistical rigor and facilitating efficient decision-making in oncology drug development.

Most recently, with the initiation of Project Optimus by the Oncology Center of Excellence (OCE) within the US Food and Drug Administration (FDA) in 2021 \citep{FDAProjectOptimus}, the exploration of efficient dose-finding strategies has become a focal point in oncology drug development. As described in the draft Guidance for Industry on dose optimization \citep{us2023optimizing} and several key publications on Project Optimus \cite{shah2021drug,blumenthal2021optimizing,fourie2022improving}, dose optimization is preferably addressed at earlier stages and before marketing approval. The challenge of dose optimization, which has been a subject of extensive investigation in clinical trials, has been explored long before the initiation of Project Optimus. Various methods have been proposed to balance the risk-benefit trade-offs, such as  EffTox \cite{thall2004dose}, BOIN12 \cite{lin2020boin12}, and those based on backfill designs \cite{dehbi2021controlled,zhao2024backfilling}. Following the introduction of Project Optimus, there has been a surge in innovative and practical methodologies aiming to integrate dose optimization with seamless design frameworks, such as the MATS design \citep{jiang2023mats}, the DROID design \citep{guo2023droid}, the controlled amplification approach \citep{dehbi2023controlled}, and the Bi3+3 design \citep{liu2023backfill} for early-phase studies. For late-phase development, integrating seamless Phase II/III design with dose optimization has emerged as an intriguing area of research. Several extensions of the original seamless design have already been proposed \citep{zhang20222,jin2022seamless,zhang2023variation,jiang_seamless_2023}. 

Existing seamless Phase II/III designs generally assess multiple doses through randomized Phase II trials. Interim analysis results typically determine whether to proceed with each arm into the Phase III part of the study. Specifically, Jin and Zhang \cite{jin2022seamless} proposed to evaluate each dose arm independently, and one or multiple dose arms may advance to Phase III if they individually demonstrate a significant treatment effect. In contrast, Zhang et al. \cite{zhang20222,zhang2023variation} recommended selecting the optimal dose at interim analysis, and then deciding whether to expand the optimal dose arm to Phase III. Jiang and Yuan \cite{jiang_seamless_2023} extend this approach by considering designs that employ different control arm enrollment strategies, aiming to address challenges in different clinical settings. Type I error control for these methods is achieved by selecting the appropriate cut-off values for test statistics when deciding whether to advance to Phase III part of the study.

%and select one or more doses for expansion to Phase III if they individually demonstrate a significant treatment effect compared to the control group, and further control Type I error based on multiple testing procedures. Designs with different control arm enrolling strategies are also discussed by \cite{jiang_seamless_2023}. 

While existing seamless Phase II/III trial designs provide valuable and practical options, several limitations exist.  First, analytically these approaches treat each dose level as an independent treatment, disregarding the potential correlation of treatment effect among dose levels, which may be readily available from Phase I dose-escalation studies (e.g., preliminary depiction of the dose-response curve). Failure to utilize this valuable prior information in a seamless design framework may incur a loss of efficiency. Moreover, the existing framework does not include a sample size re-estimation (SSR) procedure \citep{proschan2009sample, mehta2011adaptive, mehta2013adaptive, pritchett2015sample}. Incorporating SSR adds adaptiveness to the design of confirmatory trials and aligns with the recommendations in the FDA's Guidance for Industry on adaptive designs (see section V.B. of \cite{us2019adaptive}).

To address the limitations of current methods, we introduce a novel  ``\textbf{S}eamless \textbf{D}esign with \textbf{D}ose \textbf{O}ptimization'' (hereinafter ``SDDO'') framework, an adaptive two-stage Phase II/III design that incorporates both dose optimization and SSR. In Stage I, candidate dosages (treatment arms) and a control arm are evaluated in a randomized setting primarily based on a short-term surrogate endpoint, leveraging the Bayesian framework to facilitate information-borrowing from prior knowledge. After Stage I, a formal interim analysis is conducted with three objectives: (1) select the optimal dosage; (2) make decisions regarding trial early termination, Phase II continuation, or Phase III expansion; (3) SSR based on Predictive Probability of Success (PPoS) if the decision is to expand to Phase III based on all observed data up till the interim analysis. In Stage II, enrollment continues for the control arm and the selected optimal dose arm. A final analysis is then conducted when the adjusted event size is reached, combining data from both Stage I and Stage II, and using traditional statistical tests to determine the significance of treatment effect.

The remainder of this article is organized as follows. In Section 2, we first provide an overview of the proposed SDDO framework, including the design schema and model setup, clinical endpoints and decision-making procedure at different stages, SSR with PPoS, and type I error control at the final analysis. Scenarios for the comprehensive simulation studies are described and simulation results are discussed in Section 3. We conclude the article in Section 4 with some practical considerations on the SDDO framework and potential future works.

%\sout{This proposed adaptive seamless phase II/III design addresses the limitations of current methods by utilizing a Bayesian framework for the incorporation of prior knowledge and the consideration of the correlated nature between dose levels and the treatment effects. By incorporating interim analyses for decision-making and sample size re-estimation, our design has the potential to save time and cost in oncology clinical trials. Through these features, our proposed design aims to enhance efficiency and statistical power, leading to improved accuracy and effectiveness in dose optimization. Ultimately, this approach has the potential to expedite the drug development process, provide more precise dosing recommendations, and improve patient outcomes in oncology clinical trials.}

\section{Methodology}
%\sout{To develop an seamless Phase II/III design with dose optimization, we propose a SDDO framework (Seamless Design with Dose Optimization) that incorporates sample size re-estimation at the interim analysis.} 

\begin{figure}[h]
    \centering 
     \includegraphics[width=1.02\textwidth]{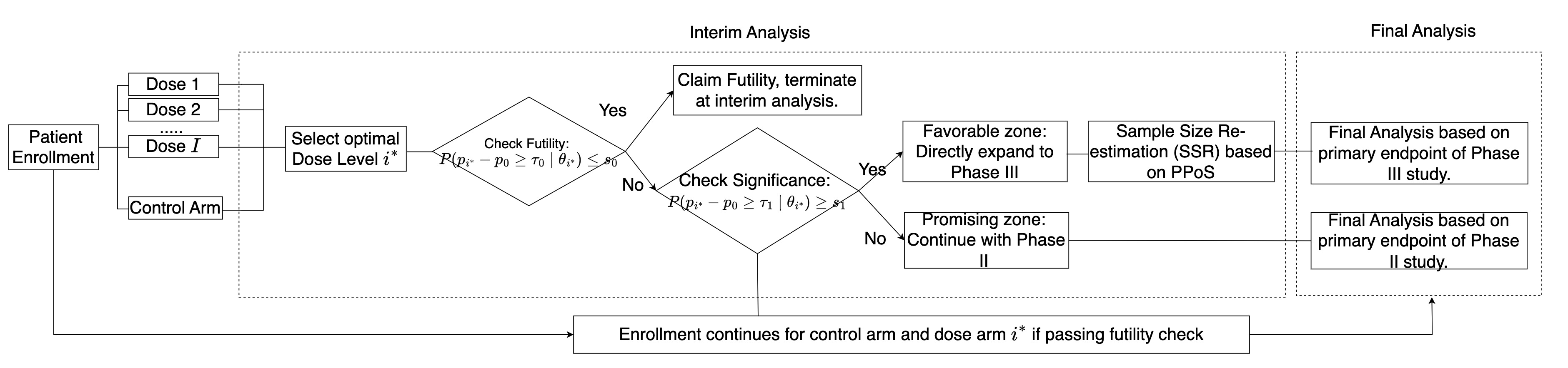}
%\label{flowchart}
\caption{Design schema of the proposed SDDO framework.}
\label{flowchart}
\end{figure}

The primary objective for the development of the SDDO framework is to enable judicious decision-making based on prior knowledge of the dose-response relationship and accumulating data evidence collected by the time of the interim analysis, while concurrently enabling more efficient sample size utilization, expediting trial timeline and maintaining statistical rigor. Figure \ref{flowchart} shows the design schema of the proposed SDDO framework.

%\vspace{-2mm}

%\subsection{Interim Analysis}
%\label{interim}
The trial begins by randomly assigning participants to either the control arm or treatment of different candidate dosages deemed worthwhile for further evaluations from early-phase studies. Interim analysis takes place when the short-term surrogate endpoint reaches sufficient level of maturity (e.g., when a pre-specified number of participants have been followed up for at least two to three tumor assessments). Subsequently, by combining data obtained from the surrogate endpoint and available (but pre-mature) data from the primary endpoint (e.g., some time-to-event endpoints accepted for regular approval) at interim analysis, we identify a single dose that demonstrates the most significant treatment effect (which we call the ``optimal dose'') \textit{among all candidate doses evaluated}. Based on this selected optimal dose, decisions regarding trial continuation are made. Specifically, three scenarios may occur:

%\sout{demonstrates the most significant treatment effect}. Based on this selected optimal dose, decisions regarding trial continuation are made. Specifically, {\color{red}{three scenarios may occur:}}

%\sout{the proposed SDDO framework primarily assesses the treatment effect using a short-term surrogate endpoint at interim analysis. The purpose of the interim analysis is three-fold: (1) Select optimal dosage; (2) Decision making on current trials; (3) Sample size adjustment for phase III expansion.}

\begin{itemize}
    \item If the optimal dose meets the ``futility'' criterion, the trial will be terminated at the interim analysis, with ``early futility'' declared at the study level.
    \item  If the optimal dose shows some treatment effects but falls short of being favorable (i.e., it lies within the ``promising zone'' but outside the ``favorable zone'' according to SSR terminology), the trial will proceed to complete Phase II, focusing solely on the selected optimal dose and maintaining the pre-determined sample size.
    \item If the interim analysis yields a favorable result (i.e., within the ``favorable zone''), the trial will directly advance to phase III with the selected optimal dose. Sample size re-estimation is conducted in this scenario to improve cost efficiency and increase Phase III success, as more investment will be committed compared to a (smaller) Phase II study.
\end{itemize}

The above decision criteria offer significant advantages and align closely with the ``quick to win, fast to fail" principle which is essential for both ethical and economic considerations in clinical trial designs. The ``quick to win'' objective is fulfilled when the interim analysis identifies the dose level in the favorable zone, enabling a swift and seamless transition to Phase III for favorable outcomes. This accelerates the development timeline and has the potential to bring effective treatments to patients more quickly. The ``fast to fail'' objective is achieved by allowing early termination of the trial when even the ``optimal dose'' turns out to be nonpromising based on interim analysis. This rapid identification of futility not only conserves valuable resources but also minimizes patient exposure to ineffective or potentially harmful treatments. Together, these features enhance cost-efficiency and align with ethical considerations, making the SDDO framework well-suited for modern clinical trials.

For many tumor types, one early surrogate endpoint commonly utilized is the overall response rate (ORR), while the primary endpoint for Phase II studies can be either ORR or progression-free survival (PFS) \citep{us2022surrogate}. In Phase III studies, the primary endpoint typically shifts to overall survival (OS) which is considered as the gold standard in oncology \cite{driscoll2009overall}. We follow this convention by using ORR as the endpoint for the interim analysis, PFS as the endpoint for Phase II, and OS as the endpoint for Phase III. It is important to note that this selection is based on the prevailing choices in oncology clinical trials, while the framework can be easily adjusted to accommodate the utilization of other endpoints as appropriate. For the rest of this paper, we assume equal randomization ratio in the control arm and each treatment arm, and denote the treatment candidate dose levels as $d_1,d_2,...,d_I$. Without loss of generality, we assume ascending dosing strength as $d_1<d_2<...<d_I$. We also denote the ORR for the $i$th dosage as $p_i$ for $i=1,2,..., I$ and the ORR for the control arm as $p_0$. Similarly, the log hazard ratio for the $i$th dosage of the treatment comparing with the control arm is denoted as $\eta_i$ for PFS, and $\theta_i$ on OS for $i=1,2,...,I$.

\subsection{Optimal Dose Selection}
\label{dose_selection}
\begin{figure}[ht]
    \centering 
     \includegraphics[width=0.8\textwidth]{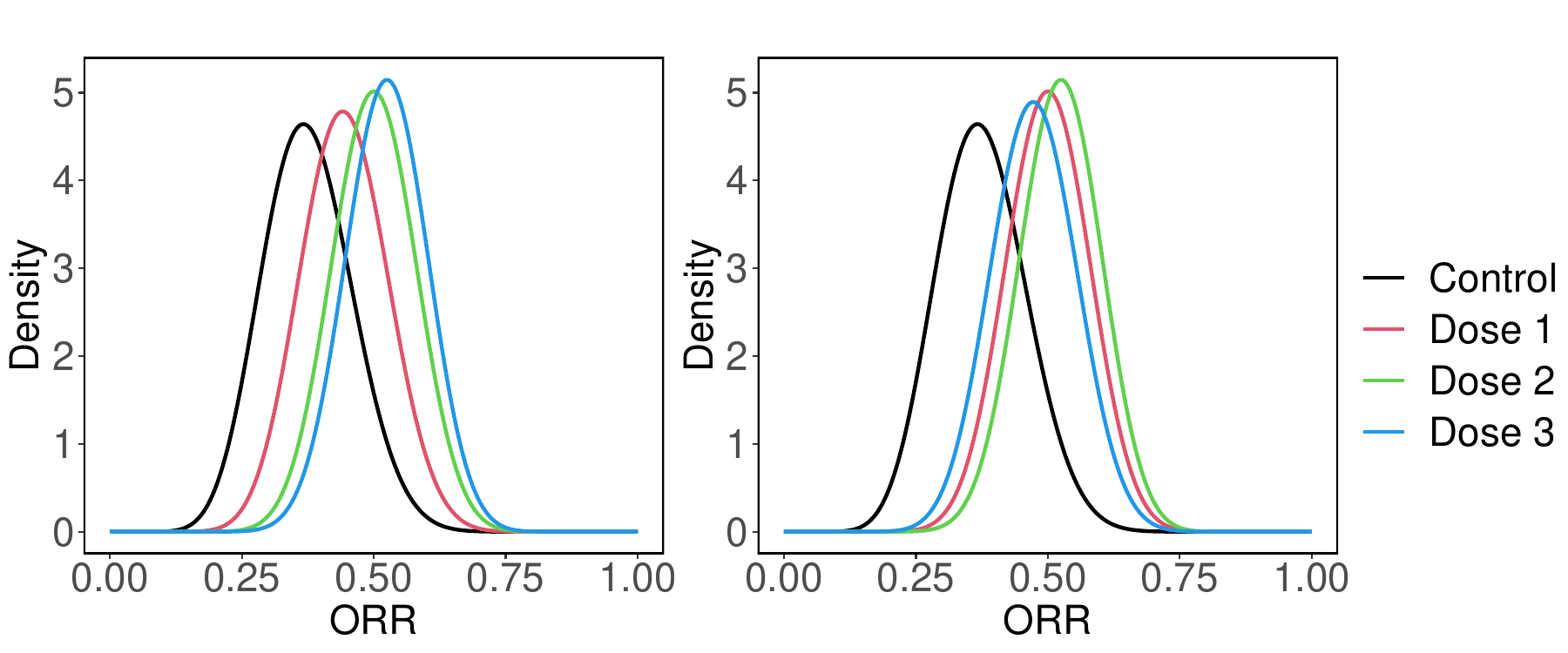}
%\label{flowchart}
\caption{Examples of prior distributions of $p_i$. In both figures, we assume $\text{Dose 1} <\text{Dose 2} <\text{Dose 3}$, and $b(d)=20$ as a constant.  $a(d)$ follows a sigmoid function in the left panel and a quadratic function in the right panel.}
\label{prior}
\end{figure}

The first objective in conducting the interim analysis is to determine the optimal dosage. To incorporate prior knowledge of the dose-response relationship, we define the prior distribution of $p_i$ as a beta distribution, denoted by $p_i\sim \text{Beta}(a(d_i),b(d_i))$, where $a(d)$ and $b(d)$ are pre-determined functions of dose level $d$. In practice, the choice of $a(d)$ and $b(d)$ can be flexible to reflect the prior knowledge (or lack thereof) of the dose-response relationship. For instance, we could specify $b(d)$ as a constant value, while $a(d)$ could be a sigmoid or quadratic function of the dose level to characterize a monotonically increasing or bell-shaped relationship, respectively. Figure \ref{prior} provides two illustrations of prior distributions across various dose arms, with the relationship implied by the relative orders of the color-coded curves. It is evident that by appropriately specifying parameter values associated with the beta distribution, we can effectively integrate prior knowledge about the dose-response relationship within the SDDO framework.

%such as $a(d_i)=a_1d_i+a_2$. In this case, a positive value of $a_1$ would indicate a positive correlation between ORR (overall response rate) and dose level, while a negative value of $a_1$ would suggest a negative correlation. 
%{\color{red}{[Note: I feel that a linear relationship may be too simplified for a typical D/R curve... should we use sigmoid curve as this example? We may also provide examples on $(a(d),b(d))$ that depict different dose-response curves, e.g., increasing and then plateued, or even a bell-shaped curve.]}}

Suppose that by the time of the interim analysis, $n_i$ participants are enrolled in the treatment arm at dose $d_i$, and $y_i$ of them are responders. We assume that the number of responders $y_i$ follows a binomial distribution $y_i|n_i,p_i \sim \text{Bin}(n_i,p_i)$. Consequently, the posterior distribution of $p_i$ can be derived as follows: 
\begin{align}
\begin{split}
f(p_i|n_i,y_i) & \propto f(n_i,y_i|p_i)f(p_i)\\
& \propto \text{Beta}(a(d_i)+y_i,b(d_i)+n_i-y_i)\\
& \approx \mathcal{N}\left(\frac{a(d_i)+y_i}{a(d_i)+b(d_i)+n_i},\frac{(a(d_i)+y_i)(b(d_i)+n_i-y_i)}{(a(d_i)+b(d_i)+n_i+1)(a(d_i)+b(d_i)+n_i)^2}\right).
\label{post_orr}
\end{split}
\end{align}

It is worth noting that during the interim analysis, a certain number of OS events may also be available. Although these observed events are likely to be limited and may not provide mature information, they can still offer additional evidence for decision-making during the interim analysis. Similar to ORR, we assume that the prior distribution of the log hazard ratio follows $\theta_i \sim \mathcal{N}(\mu(d),\frac{4}{\sigma(d)})$, where $\mu(d)$ and $\sigma(d)$ are pre-defined functions of the dose level $d_i$ that reflect prior beliefs regarding the relationship between dose levels and their corresponding hazard ratios. Based on the data collected up until the interim analysis, the data likelihood follows $\text{data}|\theta_i \sim \mathcal{N}(\log(h_i),\frac{4}{M^1_i})$, where $h_i$ represents the estimated hazard ratio based on the available data, and $M^1_i$ denotes the total number of events in the control arm and treatment arm of dose $d_i$ at interim analysis \citep{fayers1997tutorial}. Consequently, the posterior distribution at the interim analysis for the log hazard ratio of OS is given by:
\begin{align}
\begin{split}
f(\theta_i|\text{data}) & \propto f(\text{data}|\theta_i)f(\theta_i)\\
& \propto \mathcal{N}\left(\frac{\log(h_i)M^1_i+\sigma(d_i)\mu(d_i)}{M^1_i+\sigma(d_i)},\frac{4}{M^1_i+\sigma(d_i)}\right).
\end{split}
\label{post_os}
\end{align}

The correlation between ORR and log hazard ratio ($\log(\text{HR})$) has been extensively studied in clinical trials \citep{louvet2001correlation, ye2020relationship}, and it is common practice to specify a pre-determined correlation coefficient $\rho_1$ between ORR and HR in a given clinical trial based on prior knowledge and external data \citep{chen20182}. Based on equations \eqref{post_orr} and \eqref{post_os}, the joint posterior distribution of $(p_i, \theta_i)$ follows a bivariate normal distribution:
\begin{align}
\begin{bmatrix}
p_i \\
\theta_i
\end{bmatrix}
\sim
\mathcal{N}\left(
\begin{bmatrix}
\mu_{p_i} \\
\mu_{\theta_i}
\end{bmatrix},
\begin{bmatrix}
\sigma_{p_i}^2 & \rho_1\sigma_{p_i}\sigma_{\theta_i} \\
\rho_1\sigma_{p_i}\sigma_{\theta_i} & \sigma_{\theta_i}^2
\end{bmatrix}
\right),
\label{post_joint}
\end{align}
where $\mu_{p_i}$, $\mu_{\theta_i}$, $\sigma_{p_i}$, and $\sigma_{\theta_i}$ are obtained by substituting the corresponding posterior means and standard deviations from equations \eqref{post_orr} and \eqref{post_os}.

The joint posterior distribution \eqref{post_joint} incorporates both prior knowledge and all available data at the interim analysis. The optimal dose level should be thus determined corresponding to the posterior distribution that is farthest away from the null distribution with mean $[\mu_{p_i},\mu_{\theta_i}]=[p_0,0]$ and estimated variance from distribution \eqref{post_joint}. Therefore, we define the optimal dosage at the interim analysis as:
\begin{equation}
i^*=\argmax_{i\in\{1,2,...,I\}}\mathbb{E}[p_i|\theta_i=\mu_{\theta_i}],
\label{opt}
\end{equation}
and the optimal dosage is then $d_{i^*}$.

It's important to note that the selection of optimal dose is primarily driven by the surrogate endpoint (i.e., ORR), as defined in equation \eqref{opt}. This choice is influenced by the fact that only ORR has matured sufficiently for interim-stage decision-making due to the relatively short follow-up time.  However, by conditioning on the posterior mean of log hazard ratio of OS, we effectively integrate supplementary information regarding the Phase III primary endpoint into the dose selection process. In particular, when ORRs among the treatment arms are comparable based on this criterion at the interim analysis, preference is given to dose levels with lower HRs that could potentially be indicative of stronger OS benefits at the final analysis.

\subsection{Decision-Making at Interim Analysis}
\label{sec:decision}
Once the optimal dose is selected based on the Bayesian framework introduced in the previous section, the next step is to determine the appropriate course of action: whether to terminate the trial at the interim analysis, continue with the originally planned Phase II trial, or directly expand to a Phase III trial with SSR.

\begin{figure}[ht]
    \centering 
     \includegraphics[width=0.9\textwidth]{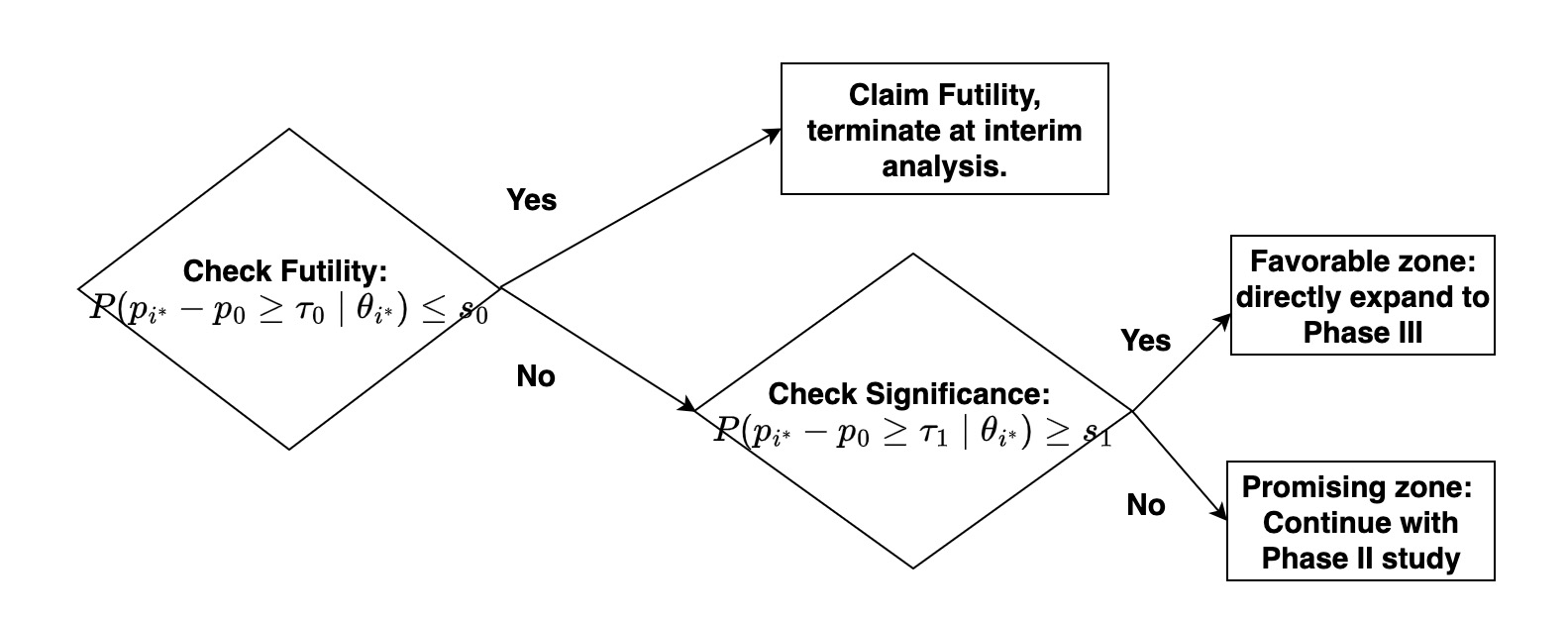}
%\label{flowchart}
\caption{Decision-making procedure for the proposed SDDO design.}
\label{decision}
\end{figure}

As depicted in Figure \ref{decision}, the decision at the interim analysis is made by evaluating futility and significance based on the joint posterior distribution of $(p_{i^*}, \theta_{i^*})$. Specifically, the selected optimal dose would be deemed futile if the probability satisfies:
\begin{equation}
P(p_{i^*} - p_0 \geq \tau_0|\theta_{i^*}) \leq s_0.
\label{futility}
\end{equation}
If the futility criterion \eqref{futility} holds, the trial would be terminated: it suggests that, although the dose itself ``outperforms'' other candidate doses, it fails to demonstrate a desired treatment effect over the control arm so that any further development is not justified. However, if the optimal dose passes the futility criterion, we proceed to assess its significance by checking:
\begin{equation}
P(p_{i^*} - p_0 \geq \tau_1|\theta_{i^*}) \geq s_1.
\label{significance}
\end{equation}
If the significance criterion \eqref{significance} holds, indicating strong evidence of a positive treatment effect at the interim analysis, the trial would directly expand to Phase III with SSR to be discussed in detail in Section \ref{ssr}. This approach saves time and sample size, as the participants already enrolled in the treatment and control arms can be further followed and included as part of the Phase III portion of the trial. %Additionally, based on the compelling interim results, it may be possible to request accelerated approval in the mean time.

%It is worth noting that if the interim results are promising, it is possible to request accelerated approval (AA) from the FDA at this stage, while continuing with the enrollment of participants. As outlined in the FDA Guidance for Industry on Accelerated Approval, the ``one-trial'' approach has been recommended as an efficient way to achieve accelerated and regular approvals within the same study. We believe that the proposed seamless design should be a good fit into this ``one-trial'' framework \citep{fashoyin2022and,fda_website}. This approach can significantly reduce the overall study duration. For instance, when equation \eqref{significance} is satisfied, along with a high PPoS as discussed in Section \ref{ssr}, there is strong evidence that the optimal dosage $d_{i^*}$ may yield positive results. Requesting accelerated approval in such cases can further expedite the timeline, as patient enrollments still continue in the meantime to collect data on OS for regular approval. Such ``one-trial'' approach could significantly reduce the time required in delivering the drug to participants.

It should be highlighted that if the interim results fall into the favorable zone, the option of seeking Accelerated Approval (AA) from the FDA becomes viable. This can be pursued concurrently with the ongoing patient enrollment process. Notably, the FDA's Guidance for Industry on AA endorses the ``one-trial" approach as an efficient means of securing both accelerated and regular approvals within the same study \citep{fashoyin2022and, FDAAA, food2023clinical}, while the proposed SDDO framework aligns well with this ``one-trial" approach. Specifically,
meeting criterion \eqref{significance} in conjunction with a high PPoS to be discussed in Section \ref{ssr} below would demonstrate compelling evidence of treatment benefits under the selected optimal dose $d_{i^*}$, and thus warrant an application for AA. Meanwhile, patient enrollment continues in this framework, with continued data collection on the primary endpoint of OS to support the pursuit of a regular approval. Our proposed SDDO framework fits into this ``one-trial" approach and has the potential to deliver novel agents to patients more efficiently.

Note that $s_0, s_1, \tau_0$, and $\tau_1$ are four crucial tuning parameters that need to be pre-determined within the SDDO framework. These parameters should be carefully selected to align with the trial's specific requirements and concerns \cite{zhao2023bayesian}, thereby maximizing the benefits of the SDDO framework. As a practical approach, we suggest setting $s_0 = s_1 = 0.9$ to ensure high confidence in decision-making. The values of $\tau_0$ and $\tau_1$ should be chosen with the aim of controlling the rate of false decisions effectively. Our approach is driven by two key considerations. Firstly, to ensure high confidence in crucial decision-making, such as deciding whether to terminate a trial or to progress to Phase III---decisions that are high-stakes and have far-reaching consequences---we recommend setting $s_0$ and $s_1$ at higher values. Secondly, considering the limited data available at interim analysis, which may obscure the true treatment effect, we propose the selection of $\tau_0$ and $\tau_1$ using grid search and simulation techniques. This approach can accommodate the inherent data variability encountered at this early stage of the trial.

From Figure \ref{decision}, given fixed values of $s_0$ and $s_1$, $\tau_0$ controls whether the trial should be terminated at the interim stage, while $\tau_1$ determines if the trial should be directly expanded to Phase III or continue with Phase II. Therefore, we propose to select $\tau_0$ by controlling the probability of falsely terminating the trial early when at least one candidate dosage shows clinically meaningful difference in ORR (i.e., a false negative decision). It is important to manage this risk of false negative decision-making, as such errors could result in missed opportunities in drug development that are of market approval potential. Meanwhile, for the value of $\tau_1$, we aim to control the rate of falsely expanding to Phase III when none of the dosage is effective, as such false positive decisions could incur substantial waste of resources and investments for Phase III study.

%\begin{figure}[ht]
%    \centering 
%     \includegraphics[width=0.6\textwidth]{tau.pdf}
%\label{flowchart}
%\caption{Mean and standard deviation of $\tau_0$ and $\tau_1$ with respect to different number of dosages based on 1000 simulations. $\tau_0$ is selected to control the false negative rate at 5\% with an ORR diffrence of 0.15. $\tau_1$ is selected to control the false positive rate at 2.5\%. The prior distribution is set to be uniform for all dosages.}
%\label{tau}
%\vspace{-3mm}
%\end{figure}

Analytically finding the value of $\tau_0$ and $\tau_1$ to achieve acceptable levels of false negative and false positive rates is computationally challenging, due to the $\max$ operator involved in \eqref{opt} for optimal dosage selection. Thus, we propose to identify the values of $\tau_0$ and $\tau_1$ by grid search between $[-1,1]$. The results via grid search are illustrated in Figure \ref{tau}. Notably, $\tau_0$ tends to remain consistent across varying numbers of candidate dosages. This stability can be attributed to the ORR difference of 0.15 being significant enough to distinguish the optimal dose arm from the control arm, and the influence of additional candidate dosages on the value of $\tau_0$ is minimal. In contrast, the value of $\tau_1$ would increase as the number of dosages increases. This finding implies the necessity to apply a more stringent multiple testing correction procedure as the number of dosages increases, for the purpose of controlling false positives.

\subsection{Sample Size Re-estimation (SSR) with PPoS}
\label{ssr}
When the decision at the interim analysis is to directly expand to Phase III, we propose to conduct sample size re-estimation (SSR) to alleviate the risk of an underpowered test at final analysis and ensure cost and time efficiency \citep{proschan2009sample, pritchett2015sample}. Following our proposed Bayesian framework, we calculate the Predictive Probability of Success (PPoS) at the interim analysis \citep{tang2015optimal, liu2020sample}, and set the new sample size as the minimum sample size required to achieve the desired PPoS.

%As per the notation introduced in Section \ref{dose_selection}, the prior distribution of the OS log hazard ratio for the optimal dosage $d_{i^*}$ is assumed to be $\theta_{i^*}\sim \mathcal{N}(\mu(d_{i^*}),\frac{4}{\sigma(d_{i^*})})$. The conditional distribution of $\hat \theta_{i^*}(t)|\theta_{i^*} \sim \mathcal{N}(\theta_{i^*},\frac{4}{M_{i^*}})$, where $\hat \theta_{i^*}(t)$ refers to the estimated log-hazard ratio at interim analysis, and $t=\frac{M^1_{i^*}}{M^1_{i^*}+\widetilde M^2}$ denotes the proportion of OS events observed at the interim analysis. Here, $\widetilde M^2$ refers to the total number of events on control arms and treatment arm $d_{i^*}$ after interim analysis. {\color{cyan} To derive the predictive distribution of $\hat \theta_{i^*}(1-t)|\theta_{i^*}(t)$, we consider the estimated log-hzarad ration of OS after the interim analysis as an} independent estimate {\color{cyan} from $\hat \theta_{i^*}(t)$ denoted as} $\hat \theta_{i^*}(1-t)|\theta_{i^*} \sim \mathcal{N}(\theta_{i^*},\frac{4}{\widetilde M^2})$, and the posterior distribution \sout{is} {\color{cyan} such as} $\theta_{i^*}|\hat \theta_{i^*}(t)\sim \mathcal{N}\left(\frac{\log(h_{i^*})M^1_{i^*}+\sigma(d_{i^*})\mu(d_{i^*})}{M^1_{i^*}+\sigma(d_{i^*})},\frac{4}{M^1_{i^*}+\sigma(d_{i^*})}\right)$. 
%\sout{To derive the predictive distribution of $\hat \theta_{i^*}(1-t)|\theta_{i^*}(t)$, we have} {\color{cyan}It follows that} 

As per the notations introduced in Section \ref{dose_selection}, the prior distribution of the OS log hazard ratio for the optimal dosage $d_{i^*}$ is assumed to be $\theta_{i^*}\sim \mathcal{N}(\mu(d_{i^*}),\frac{4}{\sigma(d_{i^*})})$. The conditional distribution of $\theta_{i^*}(t)$ and $\theta_{i^*}(1-t)$ given $\theta_{i^*}$ follows $\hat \theta_{i^*}(t)|\theta_{i^*} \sim \mathcal{N}(\theta_{i^*},\frac{4}{M_{i^*}^1})$ and  $\hat \theta_{i^*}(1-t)|\theta_{i^*} \sim \mathcal{N}(\theta_{i^*},\frac{4}{\widetilde M^2})$, respectively, where  $t=\frac{M^1_{i^*}}{M^1_{i^*}+\widetilde M^2}$ denotes the proportion of OS events observed at the interim analysis, $\hat \theta_{i^*}(t)$ and $\hat \theta_{i^*}(1-t)$ refer to the estimated log-hazard ratios based on data collected prior to and after interim analysis, respectively, and $\widetilde M^2$ refers to the total number of events on the control arm and treatment arm $d_{i^*}$ after interim analysis. The posterior distribution of $\theta_{i^*}$ after interim analysis follows $\theta_{i^*}|\hat \theta_{i^*}(t)\sim \mathcal{N}\left(\frac{\log(h_{i^*})M^1_{i^*}+\sigma(d_{i^*})\mu(d_{i^*})}{M^1_{i^*}+\sigma(d_{i^*})},\frac{4}{M^1_{i^*}+\sigma(d_{i^*})}\right)$. 

To derive the predictive distribution of $\hat \theta_{i^*}(1-t)|\theta_{i^*}(t)$, we have: 
\begin{align}
\begin{split}
E\Big(\hat{\theta}_{i^*}(1-t) \mid \hat{\theta}_{i^*}(t)\Big)
&=E\Big(E(\hat{\theta}_{i^*}(1-t) \mid \hat{\theta}_{i^*}(t), \theta_{i^*})\Big)\\
& =E \Big(\theta_{i^*} \mid \hat{\theta}_{i^*}(t)\Big) \\
&=\frac{\log(h_{i^*})M^1_{i^*}+\sigma(d_{i^*})\mu(d_{i^*})}{M^1_{i^*}+\sigma(d_{i^*})}, \\
\end{split}
\label{pred_mean}
\end{align}
\vspace{-4mm}
\begin{align}
\begin{split}
\operatorname{Var}\Big(\hat{\theta}_{i^*}(1-t) \mid \hat{\theta}_{i^*}(t)\Big)& =E\Big(\operatorname{Var}(\hat{\theta}_{i^*}(1-t) \mid \theta_{i^*}, \hat{\theta}_{i^*}(t))\Big)+\operatorname{Var}\Big(E(\hat{\theta}_{i^*}(1-t) \mid \theta_{i^*}, \hat{\theta}_{i^*}(t))\Big) \\
& = \frac{4}{\widetilde M^2}+\operatorname{Var}\Big(\theta_{i^*} \mid \hat{\theta}_{i^*}(t)\Big) =\frac{4}{\widetilde M^2}+\frac{4}{M^1_{i^*}+\sigma(d_{i^*})}.
\end{split}
\label{pred_var}
\end{align}
Based on \eqref{pred_mean} and \eqref{pred_var}, the predictive distribution is then: 
\begin{equation}
\hat \theta_{i^*}(1-t)|\hat \theta_{i^*}(t) \sim \mathcal{N}\Big(\frac{\log(h_{i^*})M^1_{i^*}+\sigma(d_{i^*})\mu(d_{i^*})}{M^1_{i^*}+\sigma(d_{i^*})},\frac{4}{\widetilde M^2}+\frac{4}{M^1_{i^*}+\sigma(d_{i^*})}\Big).
\end{equation}

At final analysis (i.e., $t=1$), the trial is considered a success if the standardized test statistic $Z(t)=Z(1)=\frac{\hat\theta_{i^*}(1)}{se(\hat \theta_{i^*}(1))}\leq z_{\alpha}$, where $\alpha$ is the desired level of Type I error rate, and then $\text{PPoS} = P(Z(1)\leq z_{\alpha}|\hat \theta_{i^*}(t))$. From \cite{kundu2023review}, we have $Z(1)=\sqrt{t}Z(t)+\sqrt{1-t}Z(1-t)$. Therefore,
\begin{align}
\begin{split}
 \text{PPoS} (\widetilde {M^2}) & = P(Z(1)\leq z_{\alpha}|\hat \theta_{i^*}(t))\\
 & = P(\sqrt{t}Z(t)+\sqrt{1-t}Z(1-t) \leq z_{\alpha}|\hat \theta_{i^*}(t)) \\
 & = P\Big(Z(1-t)\leq \frac{z_{\alpha}-\sqrt{t}Z(t)}{\sqrt{1-t}} \Big| \hat \theta_{i^*}(t)\Big)\\
 & = P\Big( Z(1-t) \sqrt{\frac{4}{\widetilde M^2}}\leq  \frac{z_{\alpha}-\sqrt{t}Z(t)}{\sqrt{1-t}} \sqrt{\frac{4}{\widetilde M^2}}\Big| \hat \theta_{i^*}(t)\Big)\\
 & = P\Big(\hat \theta_{i^*}(1-t)\leq \frac{z_{\alpha}-\sqrt{t}Z(t)}{\sqrt{1-t}} \sqrt{\frac{4}{\widetilde M^2}}\Big| \hat \theta_{i^*}(t)\Big)\\
 & = \Phi\left( \frac{\frac{2\big(z_{\alpha}-\sqrt{t}Z(t)\big)}{\sqrt{(1-t)\widetilde M^2}}-\frac{\log(h_{i^*})M^1_{i^*}+\sigma(d_{i^*})\mu(d_{i^*})}{M^1_{i^*}+\sigma(d_{i^*})}}{\sqrt{\frac{4}{\widetilde M^2}+\frac{4}{M^1_{i^*}+\sigma(d_{i^*})}}}  \right).
 \end{split}
\label{ppos}
\end{align}

The analytical solution of PPoS is given by equation \eqref{ppos}. It is important to note that all terms in equation \eqref{ppos} are fixed at interim analysis, except for $\widetilde{M^2}$ and $t$. Consequently, PPoS can be considered as a function of $\widetilde{M^2}$. To determine the adjusted sample size for the Phase III study, we define:
\begin{equation}
\widetilde {M^2}^*=\argmin_{\widetilde {M^2}}\{\text{PPoS}(\widetilde {M^2})\geq 1-\beta\},
\label{m2_star}
\end{equation}
where $1-\beta$ is the desired statistical power.

Solving equation \eqref{m2_star} explicitly is computationally challenging due to the complex analytical form in equation \eqref{ppos} and the non-monotonic nature of PPoS with respect to $\widetilde{M^2}$ \citep{chen2019application}. Furthermore, in clinical trials, event sizes are generally only expected to increase \citep{us2019adaptive}, and typically there is an upper limit on the number of subjects to be enrolled, denoted as $M^2_{\text{max}}$, to ensure the budget is under control \citep{jin2022seamless}. Consequently, the adjusted sample size denoted as $\widehat{M^2}^*$ should fall within the range $\widehat{M^2}^* \in [M^2, M^2_{\text{max}}]$, where $M^2$ denotes the pre-determined Phase III event size (without SSR).

Taking these practical considerations into account, we propose a numerical approach to solve for $\widehat{M^2}^*$ as follows.  First, we check if $\text{PPoS}(M^2) \geq 1 - \beta$. If this condition is satisfied, the event size remains the same as pre-determined $M^2$, and no sample size re-estimation is needed.  However, if $\text{PPoS}(M^2) < 1 - \beta$, we perform a grid search for $\widehat{M^2}^*$ within the range $[M^2, M^2_{\text{max}}]$ to find the smallest value that achieves the desired PPoS. If such a value exists, we take it as the adjusted event size $\widehat{M^2}^*$; if no value within the range satisfies the desired PPoS, we set $\widehat{M^2}^* = M^2_{\text{max}}$. The adjusted sample size obtained from the  approach above could be summarized as follows:
%\begin{align*}
%\widehat {M^2}^* = 
%\begin{cases}
%    M^2, & \text{if } \widetilde {M^2}^* < M^2; \\
%    \widetilde{M^2}^*, & \text{if } \widetilde {M^2}^* \in [M^2,M^2_{max}]; \\
%    M^2_{max}, & \text{if } \widetilde {M^2}^* >M^2_{max}.
%\end{cases}
%\end{align*}

\begin{align*}
\widehat{M^2}^* = 
\begin{cases}
    \argmin\limits_{\widetilde{M^2} \in [M^2, M^2_{\text{max}}]} \text{PPoS}(\widetilde{M^2}) \geq 1-\beta, & \text{if } \max\limits_{\widetilde{M^2} \in [M^2, M^2_{\text{max}}]} \text{PPoS}(\widetilde{M^2}) \geq 1-\beta, \\
    M^2_{\text{max}}, & \text{otherwise}.
\end{cases}
\end{align*}
By adopting this numerical approach and confining our grid search within the range $[M^2, M^2_{\text{max}}]$, we significantly reduce the computational resources required relative to solving equation \eqref{m2_star} directly.

\subsection{Final Analysis and Type I Error Control}
Previous sections provide a comprehensive discussion of the objectives and analyses conducted at the interim stage for the SDDO framework, which can be summarized into three components: (1) optimal dose selection, (2) expansion decision, and the possibility of requesting an accelerated approval, and (3) SSR with PPoS. Based on the decision made at the interim analysis, we proceed to enroll participants with the adjusted sample size for the final analysis. Specifically,

\begin{itemize}
    \item If the decision at the interim analysis is to declare futility, the trial will be terminated.
    \item If the decision is to continue with Phase II, the trial will continue patient enrollment in both the control and optimal dose arms until reaching the pre-determined total event size for Phase II study. Once the pre-planned PFS event size is met, a log-rank test will be performed using the combined PFS data from before and after the interim analysis to assess the significance of treatment effects at the conclusion of the Phase II study.
    \item In the case of expanding to Phase III, enrollment also continues with the control arm and the optimal dose arm until the adjusted event size $M^1_{i^*}+\widehat {M^2}^*$ for OS is achieved. Subsequently, a log-rank test is conducted based on all the available OS data from the beginning of trial to assess the significance of treatment effects.
\end{itemize}

Controlling Type I error at final analysis is a critical aspect in seamless designs \citep{chen20182,zhang2023variation}. In our proposed SDDO framework, there are two principal sources that may contribute to the inflation of Type I error:
\begin{enumerate}
\item \textbf{Multiple Testing Issue}: The process of selecting the optimal dose from multiple candidate dosages during interim analysis introduces the issue of multiple testing. 
\item \textbf{Interim Analysis and Phase III expansion}:  Inclusion of data from the primary endpoint for the Phase III study in the decision-making process at the interim analysis stage can also incur Type I error inflation at the final analysis.
\end{enumerate}

In the proposed SDDO framework, the two sources of Type I error inflation are addressed through the careful selection of $\tau_0$ (for futility) and $\tau_1$ (for significance). Specifically, the choice of $\tau_0$ results in the early termination of a certain proportion of trials during the interim analysis, thereby precluding their advancement to the final analysis stage. This proportion of terminated trials would never have the opportunity to detect any potential positive treatment effects in the final analysis. Consequently, $\tau_0$ serves dual purposes: beyond controlling the probability of erroneously terminating the trial, it also functions as a ``correction mechanism'' to the multiple testing issue. This addresses the first source of overall Type I error inflation.

In parallel, the selection of \(\tau_1\) also serves dual purposes: it not only restricts the probability of incorrect expansion to Phase III when none of the treatment arms has demonstrated efficacy but also manages the probability of directly expanding to Phase III. By constraining the probability of advancing to Phase III, \(\tau_1\) effectively controls the second source of Type I error inflation.

While theoretical guarantees for Type I error control are challenging to achieve when Bayesian methods are applied, we provide a theoretical proof for Type I error control, assuming uniform priors, in Appendix Section \ref{sec:type1}. Extensive simulations, to be elaborated in Section \ref{simu}, affirm that the proposed SDDO framework succeeds in controlling Type I error rates while preserving statistical powers.

\section{Simulations}
\label{simu}

To evaluate our proposed SDDO framework, we conduct extensive simulation studies to assess its operating characteristics in a wide spectrum of realistic and representative scenarios. In all simulation settings, we initiate the Phase~II study with three candidate dosages and assume an equal randomization ratio between each treatment arm and the control arm. An interim analysis is performed after 60 participants in each arm have reached mature ORR, and at least 30 OS events have been observed in the control arm. Unless otherwise stated, we assume the correlation between the surrogate endpoint (ORR) and the primary endpoint (OS) for the Phase~III study (using \( \log(\text{HR}_{\text{OS}}) \)) to be \( \rho_1 = -0.5 \).

The total event size for the Phase~II study is set at 140, with the aim of detecting a hazard ratio of \( 0.58 \) with \( 90\% \) power and a one-sided Type~I error rate of \( 0.025 \). For the pre-planned Phase~III study, the event size is set as \( M^2 = 226 \), and the maximum event size for Phase~III is \( M^2_{\max} = 507 \). These settings are designed to detect hazard ratios of \( 0.65 \) and \( 0.75 \), respectively, with \( 90\% \) power and a one-sided type~I error rate of \( 0.025 \). For tuning parameters, we set \( s_0 = s_1 = 0.9 \), and select \( \tau_0 \) and \( \tau_1 \) by grid search to control the false positive rate at 2.5\% and false negative rate at 5\% as described in Section~\ref{sec:decision}. This results in \( \tau_0 = -0.05 \) and \( \tau_1 = 0.10 \). 

%For all simulation settings in Table \ref{standard}, \ref{hold} and \ref{violate}, the first entry in ORR, $\text{HR}_{\text{PFS}}$ and $\text{HR}_{\text{OS}}$ refer to the control arm, and the other entries refers to three different dosages, assuming $d_1<d_2<d_3$. The column \textit{Optimal Dose Percentage} refers to the probability of selecting $d_1,d_2,d_3$ as the optimal dosage at interim analysis.  Uninformative priors are assigned to ORR and  $ \log(\text{HR}_{\text{OS}})$  for all simulation settings. Specifically, the prior distribution for ORR is $\text{Beta}(2,2)$, and $\mathcal{N}(0,0.5)$ for $ \log(\text{HR}_{\text{OS}})$ across all dosages. The reported expected event size only include events occurred in the selected dose arm at interim analysis and the control arm.

In Table \ref{standard}, \ref{hold}, and \ref{violate}, the first entries under ORR, \(\text{HR}_{\text{PFS}}\), and \(\text{HR}_{\text{OS}}\) correspond to the control arm, while the subsequent entries refer to three candidate dosages, assuming \(d_1 < d_2 < d_3\). The column labeled \textit{Optimal Dose Percentage} demonstrates the probability of selecting \(d_1, d_2,\) or \(d_3\) as the optimal dosage during the interim analysis, respectively. Uninformative priors are assigned to ORR and \(\log(\text{HR}_{\text{OS}})\) across all simulation settings in Table \ref{standard}-\ref{violate}. Specifically, the prior distribution for ORR is modeled as \(\text{Beta}(2,2)\), and \(\mathcal{N}(0,0.5)\) for \(\log(\text{HR}_{\text{OS}})\) across all dosage levels. It should be noted that the reported expected event sizes include only the events that occurred in the optimal dose arm selected at the interim analysis and the control arm. Simulation results using different prior distributions are presented in Tables \ref{prior_simu} and \ref{weak_prior_simu}.

\begin{table}[t]
\centering
\footnotesize % Make the font size smaller
\setlength{\tabcolsep}{2pt} % Reduce the horizontal padding between columns
\begin{tabular}{|l|c|c|c|c|c|c|}
\hline
\multicolumn{1}{|c|}{\begin{tabular}[c]{@{}c@{}}Simulation \\ Scenarios\end{tabular}} & \begin{tabular}[c]{@{}c@{}}Optimal Dose \\ Percentage \end{tabular} & & \begin{tabular}[c]{@{}c@{}}Interim \\ Decision\end{tabular} & \begin{tabular}[c]{@{}c@{}}Positive \\ Rate\end{tabular} & \begin{tabular}[c]{@{}c@{}}Expected \\ Event Size\end{tabular} & \begin{tabular}[c]{@{}c@{}}Expected \\ Study Duration\end{tabular} \\
\hline
\multirow{4}{*}{\begin{tabular}[c]{@{}l@{}}Global Null:\\ ORR = {[}0.2,0.2,0.2,0.2{]}\\ $\text{HR}_{\text{PFS}}$={[}1,1,1,1{]}\\ $\text{HR}_{\text{OS}}$={[}1,1,1,1{]}\end{tabular}} & \multirow{4}{*}{\footnotesize{[33.32\%,33.35\%,33.33\%]}} & Overall & -- & 1.45\% & 120 & 23.57 \\
\cline{3-7}
& & Terminate & 44.12\% & 0 & 79 & 17.63 \\
\cline{3-7}
& & Phase II & 53.69\% & 2.37\% & 140 & 27.89 \\
\cline{3-7}
& & Phase III & 2.19\% & 8.22\% & 453 & 37.15 \\
\hline
\multirow{4}{*}{\begin{tabular}[c]{@{}l@{}}One Significant:\\ ORR = {[}0.2,0.2,0.2,0.35{]}\\ $\text{HR}_{\text{PFS}}$={[}1,1,1,0.58{]}\\ $\text{HR}_{\text{OS}}$={[}1,1,1,0.7{]}\end{tabular}} & \multirow{4}{*}{\footnotesize{[1.11\%,0.98\%,97.91\%]}} & Overall & -- & 88.67\% (88.65\%) & 280 & 38.35 \\
\cline{3-7}
& & Terminate & 3.11\% & 0 & 74 & 18.50 \\
\cline{3-7}
& & Phase II & 44.88\% & 86.01\% (85.98\%) & 140 & 36.01 \\
\cline{3-7}
& & Phase III & 52.01\% & 96.27\% (96.25\%) & 414 & 42.67 \\
\hline
\multirow{4}{*}{\begin{tabular}[c]{@{}l@{}}Two Significants:\\ ORR = {[}0.2,0.2,0.35,0.35{]}\\ $\text{HR}_{\text{PFS}}$={[}1,1,0.58,0.58{]}\\ $\text{HR}_{\text{OS}}$={[}1,1,0.7,0.7{]}\end{tabular}} & \multirow{4}{*}{\footnotesize{[0.16\%,50.88\%,48.96\%]}} & Overall & -- & 93.98\% (93.98\%) & 327 & 40.67 \\
\cline{3-7}
& & Terminate & 0.66\% & 0 & 74 & 19.27 \\
\cline{3-7}
& & Phase II & 29.45\% & 89.24\% (89.24\%) & 140 & 36.30 \\
\cline{3-7}
& & Phase III & 69.89\% & 96.85\% (96.85\%) & 408 & 42.72 \\
\hline
\multirow{4}{*}{\begin{tabular}[c]{@{}l@{}}Global Alternative:\\ ORR = {[}0.2,0.35,0.35,0.35{]}\\ $\text{HR}_{\text{PFS}}$={[}1,0.58,0.58,0.58{]}\\ $\text{HR}_{\text{OS}}$={[}1,0.58,0.58,0.58{]}\end{tabular}} & \multirow{4}{*}{\footnotesize{[32.54\%,34.03\%,33.43\%]}} & Overall & -- & 98.61\% (98.61\%) & 302 & 42.81 \\
\cline{3-7}
& & Terminate & 0.05\% & 0 & 68 & 19.73 \\
\cline{3-7}
& & Phase II & 11.94\% & 89.86\% (89.86\%) & 140 & 36.40 \\
\cline{3-7}
& & Phase III & 88.01\% & 99.85\% (99.85\%) & 324 & 43.69 \\
\hline
\end{tabular}
\caption{Operating characteristics of the SDDO framework based on 10,000 simulation runs under global null, only one significant dosage, two significant dosages, and global alternative scenarios. ``Positive Rate" column shows overall positive effect rate, and in brackets, the rate of selecting the optimal dose with a positive effect.} 
\label{standard}
\end{table}

Table \ref{standard} presents the operating characteristics under various standard scenarios. Under the global null hypothesis, where no candidate dosage demonstrates treatment effect, the proposed SDDO framework successfully controls the rate of false entry into Phase III trials at \(2.19\%\). This is achieved by selecting the appropriate value of \(\tau_1\) through grid search. Additionally, the overall Type I error rate is well-controlled at \(1.45\%\). In the second scenario with only one candidate dosage showing treatment effect, the SDDO framework correctly identifies the optimal dose with a \(97.91\%\) probability. Moreover, the chance of prematurely terminating the trial is capped at \(3.11\%\), thanks to the tuning parameter \(\tau_0\). In the third scenario when two dosages show treatment effect, they are almost equally selected as the optimal dose, with similar operating characteristics as seen in the "one significant" scenario. Under the global alternative hypothesis with all candidate dosages showing treatment effects, the proposed framework exhibits strong statistical power, achieving a true positive rate of \(98.61\%\). Furthermore, the expected event size for Phase III can be significantly reduced (i.e., 324 instead of 453, 414, and 408 in the other three scenarios), as such a strong signal allows for effect detection with fewer events. For the overall development timeline (study duration), in case that none of the dosages is effective, the study on average can be completed within two years (23.57 months), while in case that at least one dosage is effective, the time from the first patient into the final Phase III OS readout readily for regular approval is around 43 months (and an accelerated approval request based on PPoS would likely be made at the earlier interim analysis stage). 

\begin{table}[t]
\centering
\footnotesize % Make the font size smaller
\setlength{\tabcolsep}{2pt}
\begin{tabular}{|l|c|c|c|c|c|c|}
\hline
\multicolumn{1}{|c|}{\begin{tabular}[c]{@{}c@{}}Simulation \\ Scenarios\end{tabular}} & \begin{tabular}[c]{@{}c@{}}Optimal Dose\\ Percentage\end{tabular} &  & \begin{tabular}[c]{@{}c@{}}Interim \\ Decision\end{tabular} & \begin{tabular}[c]{@{}c@{}}Positive \\ Rate\end{tabular} & \begin{tabular}[c]{@{}c@{}}Expected \\ Event Size\end{tabular} & \begin{tabular}[c]{@{}c@{}}Expected \\ Study Duration\end{tabular} \\ \hline
\multirow{4}{*}{\begin{tabular}[c]{@{}l@{}}Same ORR, Different HR:\\ ORR = {[}0.2,0.35,0.35,0.35{]}\\ $\text{HR}_{\text{PFS}}$={[}1,0.8,0.7,0.6{]}\\ $\text{HR}_{\text{OS}}$={[}1,0.8,0.7,0.6{]}\end{tabular}} & \multirow{4}{*}{{[}20.48\%,32.25\%,47.27\%{]}} & Overall & -- & 86.96\% (44.72\%) & 332 & 41.02 \\ \cline{3-7} 
 &  & Terminate & 0.22\% & 0 & 73 & 19.53 \\ \cline{3-7} 
 &  & Phase II & 20.01\% & 61.37\% (35.32\%) & 140 & 33.83 \\ \cline{3-7} 
 &  & Phase III & 79.77\% & 93.62\% (47.20\%) & 380 & 32.89 \\ \hline
\multirow{4}{*}{\begin{tabular}[c]{@{}l@{}}Weak Correlation, \footnotesize{$\rho_1=-0.1$}:\\ ORR = {[}0.2,0.35,0.35,0.35{]}\\ $\text{HR}_{\text{PFS}}$={[}1,0.8,0.7,0.6{]}\\ $\text{HR}_{\text{OS}}$={[}1,0.8,0.7,0.6{]}\\ \end{tabular}} & \multirow{4}{*}{{[}29.70\%,33.65\%,36.65\%{]}} & Overall & -- & 71.86\% (32.99\%) & 274 & 38.02 \\ \cline{3-7} 
 &  & Terminate & 1.43\% & 0 & 69 & 17.47 \\ \cline{3-7} 
 &  & Phase II & 49.72\% & 56.21\%  (30.37\%)& 140 & 33.46 \\ \cline{3-7} 
 &  & Phase III & 48.85\% & 89.89\% (36.62\%)& 416 & 43.27 \\ \hline
\multirow{4}{*}{\begin{tabular}[c]{@{}l@{}}Weak Effect:\\ ORR = {[}0.2,0.25,0.27,0.3{]}\\ $\text{HR}_{\text{PFS}}$={[}1,0.95,0.9,0.85{]}\\ $\text{HR}_{\text{OS}}$={[}1,0.95,0.9,0.85{]}\end{tabular}} & \multirow{4}{*}{{[}14.06\%,29.20\%,56.84\%{]}} & Overall & -- & 19.13\% (14.91\%) & 221 & 31.69 \\ \cline{3-7} 
 &  & Terminate & 7.44\% & 0 & 78 & 18.49 \\ \cline{3-7} 
 &  & Phase II & 66.03\% & 11.83\% (8.79\%) & 140 & 29.73 \\ \cline{3-7} 
 &  & Phase III & 26.53\% & 42.67\% (34.31\%) & 462 & 40.26 \\ \hline
\end{tabular}
\caption{Operating characteristics of the SDDO framework based on 10,000 simulation runs under various scenarios assuming that a higher ORR can translate into benefits in PFS and OS. ``Positive Rate" column shows overall positive effect rate, and in brackets, the rate of selecting the optimal dose with a positive effect.}
\label{hold}
\end{table}

Table \ref{hold} presents simulation results under a range of realistic settings other than the standard settings, assuming that a higher ORR can translate into benefits in PFS and OS. In the first setting, we consider a scenario where all dose levels yield comparable ORR but differ in PFS and OS benefits. By conditioning the optimal dose selection at interim analysis on the OS data, we achieve a higher probability (47.27\%) of selecting the true optimal dose that maximizes OS benefits while attaining a robust statistical power of 86.96\%. However, when only the correlation between the surrogate and primary endpoints is changed to be lower, as illustrated in the second setting (\(\rho_1 = -0.1\) compared to \(\rho_1 = -0.5\) in other settings), both the likelihood of identifying the true optimal dose (36.65\% vs. 47.27\%) and the overall statistical power (71.86\% vs. 86.96\%) are diminished. This decline occurs because a lower correlation \(\rho_1\) results in a lower weighting of OS benefits in the optimal dose selection, which further affects the overall statistical power. This observation strongly suggests that choosing a surrogate endpoint with a higher correlation to the primary endpoint can enhance the performance of the proposed SDDO framework. Lastly, we examine scenarios in which all dosages demonstrate a weak treatment effect. Our results indicate that, while the SDDO framework could correctly select the true optimal dose with a higher probability (56.84\%), in most cases, it fails to detect a significant treatment effect in the final analysis as expected (only 11.83\% and 42.67\% for PFS and OS, respectively), and the false-positive rate remains controlled at 19.13\%.

\begin{table}[t]
\centering
\footnotesize % Make the font size smaller
\setlength{\tabcolsep}{2pt}
\begin{tabular}{|l|c|c|c|c|c|c|}
\hline
\multicolumn{1}{|c|}{\begin{tabular}[c]{@{}c@{}}Simulation \\ Scenarios\end{tabular}} & \begin{tabular}[c]{@{}c@{}}Optimal Dose\\ Percentage\end{tabular} &  & \begin{tabular}[c]{@{}c@{}}Interim \\ Decision\end{tabular} & \begin{tabular}[c]{@{}c@{}}Positive \\ Rate\end{tabular} & \begin{tabular}[c]{@{}c@{}}Expected \\ Event Size\end{tabular} & \begin{tabular}[c]{@{}c@{}}Expected \\ Study Duration\end{tabular} \\ \hline
\multirow{4}{*}{\begin{tabular}[c]{@{}l@{}}Strong PFS, Weak OS:\\ ORR = {[}0.2,0.3,0.35,0.4{]}\\ $\text{HR}_{\text{PFS}}$={[}1,0.6,0.6,0.6{]}\\ $\text{HR}_{\text{OS}}$={[}1,0.95,0.9,0.85{]}\end{tabular}} & \multirow{4}{*}{{[}5.80\%,25.03\%,69.17\%{]}} & Overall & -- & 54.23\% (39.15\%) & 373 & 39.34 \\ \cline{3-7} 
 &  & Terminate & 0.75\% & 0 & 80 & 19.85 \\ \cline{3-7} 
 &  & Phase II & 30.40\% & 85.89\% (52.23\%) & 140 & 35.75 \\ \cline{3-7} 
 &  & Phase III & 68.85\% & 40.84\% (33.80\%) & 479 & 41.14 \\ \hline
\multirow{4}{*}{\begin{tabular}[c]{@{}l@{}}Weak PFS, Strong OS:\\ ORR = {[}0.2,0.3,0.35,0.4{]}\\ $\text{HR}_{\text{PFS}}$={[}1,0.95,0.9,0.85{]}\\ $\text{HR}_{\text{OS}}$={[}1,0.75,0.7,0.65{]}\end{tabular}} & \multirow{4}{*}{{[}5.44\%,25.12\%,69.44\%{]}} & Overall & -- & 84.66\% (61.00\%)& 395 & 41.13 \\ \cline{3-7} 
 &  & Terminate & 0.17\% & 0 & 74 & 20.38 \\ \cline{3-7} 
 &  & Phase II & 15.60\% & 12.11\%  (9.03\%) & 140 & 29.89 \\ \cline{3-7} 
 &  & Phase III & 84.23\% & 98.25\% (70.75\%) & 395 & 43.26 \\ \hline
\multirow{4}{*}{\begin{tabular}[c]{@{}l@{}} Strong ORR, Weak HR:\\ ORR = {[[}0.2,0.3,0.35,0.4{]}\\ $\text{HR}_{\text{PFS}}$={[}1,1,0.95,0.9{]}\\ $\text{HR}_{\text{OS}}$={[}1,1,0.95,0.9{]}\\ \end{tabular}} & \multirow{4}{*}{{[}5.82\%,25.12\%,69.06\%{]}} & Overall & -- & 15.62\% (13.31\%) & 366 & 36.41 \\ 
\cline{3-7} 
 &  & Terminate & 0.92\% & 0 & 81 & 19.60 \\ \cline{3-7} 
 &  & Phase II & 34.10\% & 7.68\% (6.04\%) & 140 & 29.06 \\ \cline{3-7} 
 &  & Phase III & 64.98\% & 20.01\% (17.31\%)& 489 & 40.50 \\ \hline
\multirow{4}{*}{\begin{tabular}[c]{@{}l@{}} Weak ORR, Strong HR:\\ ORR = {[}0.2,0.22,0.25,0.28{]}\\ $\text{HR}_{\text{PFS}}$={[}1,0.7,0.65,0.6{]}\\ $\text{HR}_{\text{OS}}$={[}1,0.7,0.65,0.6{]}\\ \end{tabular}} & \multirow{4}{*}{{[}10.55\%,29.23\%,60.22\%{]}} & Overall & -- & 83.19\% (53.49\%) & 212 & 37.39 \\ \cline{3-7} 
 &  & Terminate & 3.36\% & 0 & 78 & 18.69 \\ \cline{3-7} 
 &  & Phase II & 56.59\% & 76.75\% (46.67\%)& 140 & 34.99 \\ \cline{3-7} 
 &  & Phase III & 40.05\% & 99.27\% (67.61\%) & 326 & 42.35 \\ \hline
\end{tabular}
\caption{Operating characteristics of the SDDO framework based on 10,000 simulation runs under various scenarios when a higher ORR does not necessarily translate into benefits in PFS or OS. ``Positive Rate" column shows overall positive effect rate, and in brackets, the rate of selecting the optimal dose with a positive effect.}
\label{violate}
\end{table}
Lastly, we assess the performance of the proposed SDDO framework in scenarios where a significant treatment effect in ORR does not invariably translate to PFS or OS benefits. Such a discrepancy between the surrogate and primary endpoints might elevate the likelihood of making sub-optimal decisions during the interim analysis. Given that similar observations have emerged in certain clinical trials, it is crucial to evaluate how well the SDDO framework can manage these instances. The pertinent results under three scenarios are discussed in Table \ref{violate}. All scenarios are deliberately chosen to reflect some associated real-world examples.

The first scenario mimics the evaluations of paclitaxel plus bevacizumab as initial treatment for metastatic breast cancer where only PFS and ORR demonstrated significant treatment effects in the E2100 trial \citep{miller2007paclitaxel} but subsequent studies failed to show the investigational treatment's OS superiority over different chemotherapy regimens \citep{miles2010phase, robert2011ribbon}. Under this setting, the proposed SDDO framework has demonstrated a commendable ability to select the true optimal dose with a high probability, consistent with the results presented in Tables \ref{standard} and \ref{hold}. Nonetheless, due to the modest OS benefits in this context, the positive rate of advancing to Phase III stands at only 40.84\%. This outcome is attributable to the inconsistencies between the surrogate and primary endpoints in predicting treatment benefits that may result from multiple factors such as different cancer types and/or lines of therapies \citep{korn2018surrogate}. Therefore, on the one hand, it underscores the importance of meticulously choosing a surrogate endpoint that can reliably predict survival benefits to optimize the performance of the SDDO framework; on the other hand, in case when the surrogacy is poorly established, the SDDO framework is advantageous over conventional designs for its capacity of quickly concluding marginal OS benefits within the same study (in this case, it may be just within months instead of multiple years of conducting separate confirmatory trials, along with substantially reduced number of participants exposed to the suboptimal treatments).

The second scenario mimics the evaluations of nivolumab as a treatment for previously treated advanced renal-cell carcinoma where both ORR and OS demonstrated significant treatment effects over everolimus but median PFS was similar between the two groups \citep{motzer2015nivolumab}. Another example is sipuleucel-T as a treatment for metastatic castration-resistant prostate cancer where time to progression (TTP, a related endpoint to PFS) is similar but OS is positive \citep{kantoff2010sipuleucel}. In comparison to the first scenario, the proposed SDDO framework exhibits a higher likelihood of advancing to the Phase III study (84.23\% vs. 68.85\%). This is attributable to the integration of OS benefits during the decision-making process (at the interim) concerning trial expansion. Furthermore, the proposed SDDO framework yields a strong statistical power of \(84.66\%\) in detecting such OS benefits. This is particularly notable since it doesn't rely on any assumptions regarding the correlation between the surrogate and primary endpoints for the Phase II study (in this case, an inconsistent ORR and PFS). Such flexibility offers additional advantages for employing the SDDO framework, as in such scenarios, traditional designs would be terminated at the end of the Phase II study since no treatment effect is observed in regard to PFS benefits, while other seamless designs require assumptions on the correlation between the surrogate endpoint and the primary endpoint for the Phase II study \citep{li2022flexible,zhang20222,zhang2023variation} which may incur a detrimental effect on concluding the survival benefits in case of poor surrogacy.

The third and fourth scenarios in Table \ref{violate} evaluate the performance of the SDDO framework when treatment effects measured between ORR and time-to-event endpoints (PFS and OS) are inconsistent. Numerous Phase III failures exemplify the third scenario as strong efficacy signals measured by early surrogate endpoints do not necessarily translate into survival benefits in late-phase development. For the fourth scenario, Kok et al. \cite{kok2021tumor} revealed that relative measures of tumor response are poor surrogate endpoints for OS for many immune checkpoint inhibitors. Under these scenarios, the proposed framework is capable of selecting the true optimal dosage (69.09\% and 60.22\%) by means of strong ORR (scenario 3) or strong OS (scenario 4) at the interim. For the most challenging and bewildering scenario 3, the SDDO framework managed to control the overall positive rate at 15.62\% with a high probability of concluding no PFS or OS benefits at the end of the study. For scenario 4 typically seen in immunotherapies, despite weak ORR signals in the interim, the SDDO framework only prematurely terminated the study 3.36\% of the time, while attaining high probabilities of trial success by PFS or OS at the final analysis.

\section{Discussion}
The Oncology Center of Excellence (OCE) at FDA initiated a couple of Research and Development Projects that ``aim to advance the development and regulation of medical products for patients with cancer''  \citep{FDAOCE}. The proposed SDDO framework provides possible solutions to two of the projects, namely Project Optimus \citep{FDAProjectOptimus} and Project FrontRunner \citep{FDAProjectFrontRunner}, which we will provide more context below.

The Phase II part of the SDDO framework aims to determine the optimal dosage by randomizing multiple treatment doses (and a control arm), which is in accordance with the guiding principles of Project Optimus. One noteworthy advantage of the SDDO framework is its capacity to formally incorporate prior dose-response information obtained from Phase I studies (as illustrated in Figure~\ref{prior}). This feature is very relevant and critical in contemporary oncology development, which has evolved from the traditional chemotherapeutic agents to a wide range of novel therapies such as molecular targeted agents, immunotherapies, and cell therapies that potentially exhibit drastically different dose-response relationships. Some of them may exhibit a ``bell-shaped'' relationship, indicating ``less (dose) is more (favorable benefit-risk profile)'' (e.g., the most recent example is tarlatamab, when the lower 10-mg group showed better efficacy and safety profiles than the higher 100-mg group \cite{ahn2023tarlatamab}). The seamless incorporation of the dose optimization component with simultaneous evaluations of proof-of-concept efficacy signals (or lack of) in a randomized pivotal trial setting has made the SDDO framework a viable development option.

%\sout{Relying on the MTD from the dose escalation phase for the recommended dose for phase 2 with assumptions that dose activity and dose toxicity relationships are strictly dose-dependent may not fit for current therapies which often have significantly less toxicity and the toxicity-activity is less correlated as chemotherapies. Thus the MTD may not necessarily be the optimal dose \citep{shord2023us}. In our SDDO framework, the randomized phase II part with multiple treatment arms for different dose levels and a concurrent control arm allows investigators to determine the optimal dose for desired treatment effects. This stage of design aligns with the outlined guidance of Project Optimus \citep{FDAProjectOptimus} to compare multiple dosages that a randomized dose-response trial prior to the large-scale study for registrational purpose to select the optimal dose. }

The proposed SDDO design is also closely related to Project Frontrunner which encourages evaluations earlier in the course of the disease (e.g., first and second lines) and FDA's accelerated approval pathway \citep{fashoyin2022and, FDAAA}. By utilizing a single randomized controlled trial (RCT) for both accelerated and regular approvals instead of two trials (conventionally involving a single-arm trial for AA and a separate trial for regular approval), the SDDO framework fits into the ``one-trial" approach \citep{fashoyin2022and,chen2023framing} with enhancements to achieve multiple objectives, namely the optimal dose selection and potential AA at the interim analysis via the use of (mature) early surrogate (e.g., ORR) and (supplementary) long-term time-to-event endpoints (e.g., OS) within a Bayesian framework; and potential regular approval at the final analysis by comparing the selected optimized dosing arm with the concurrent control arm, with possible SSR to increase the trial's success rate. The many advantages brought by the SDDO framework, such as shortening the \textit{overall} study duration and making the best use of patient resources, are expected to greatly increase the efficiency of oncology development.

%\sout{which has sample size powered by long term endpoint e.g. PFS or OS information incorprated simutaneously at interim analysis via Bayesian approach to detect clinically meangingful and statistically significant improvement of the treatment arm with optimized dose over the control arm. Such feature can significantly reduce overall study duration because it allows investigators to quickly move to accelerate approval followed by a Phase III study with concurrent control from the first stage and powered by SSR if the treatment benefits patients most and terminate immediately if declared futility.}

%\sout{We propose ORR as the surrogate endpoint for the interim analysis because it is not only a commonly used endpoint for accelerated approval depending on the context of use \citep{us2022surrogate} but also usually can mature quicker, two or three tumor assessments with median scan frequency of every 8 weeks \citep{haslam2022frequency}, than time-to-event endpoints such as PFS or OS.} 

We propose ORR as the surrogate endpoint for the interim analysis (dose optimization and preliminary efficacy evaluations) because it matures quicker (e.g., two or three tumor assessments with median scan frequency of every 8 weeks \citep{haslam2022frequency}) than time-to-event endpoints such as PFS and OS, and had been accepted in many cases for accelerated approvals depending on the context of use \citep{us2022surrogate}. Slight modifications of the SDDO framework would allow the use of PFS for such decision-making (especially, when PFS is a better predictor of OS than ORR). It is worth emphasizing that the choice of an appropriate surrogate endpoint requires rigorous validations in terms of its clinical relevance to the final primary endpoint, whether its improvement can be generally ``translated'' into meaningful improvement as measured by the primary endpoint, and consistency for the relationship between the surrogate and final endpoints in the particular population of interest. We evaluated some challenging but realistic scenarios when surrogacy is less established (see Table \ref{violate}): unsurprisingly, the performance of the SDDO framework is undermined relative to those scenarios with good surrogacy. However, for conflicting cases between ORR and OS (scenarios 3 and 4), the SDDO framework is still preferred over traditional (two-trial) designs in that it can either ``stop loss'' much earlier (scenario 3, when a promising ORR does not translate into survival benefits), or give a ``second chance'' to those investigational agents which actual survival benefits could not be predicted by an early surrogate endpoint (scenario 4). In each case, a traditional design would likely require a separate Phase III study to demonstrate the lack of survival benefits (scenario 3), or would make an incorrect, pre-mature termination decision (scenario 4).

%\sout{The shorter turnaround time can faster clinical trial duration then reduce the cost; be more ethical for patients to not wait for long term endpoint for evidence of benefit; increased participantion rate results in faster recruitment and retention; expedited drug approval, etc. Even if PFS can be adopted to SDDO framework at interim analysis, ORR can be more suitable for dose optimization purpose since the efficacy data of the dose level candidates from dose escalation phase can be preliminary due to short follow up time and small samples.} 

%\sout{predictivity of the changes in surrogate can be translatable to the changes in final primary endpoint,}

%\sout{but emphasizes the importance of the choice of appropriate surrogate endpoint. Especially, in cases when there is weak ORR benefit of the treatment arms but trial eventually shows survival benefit, use of SDDO framework is discouraged. Even though simulation results show a relatively high overall positive rate, such discrepancy would make the decision of not terminating trial at the interim analysis too risky and possibly result in excessive time or monetary costs if the weak benefit in ORR is truly not translatable to survival benefit. }

The SDDO framework may encounter some operational and logistical challenges. The adaptation decision is made when the surrogate endpoint matures with a certain fraction of OS events observed in the control arm. Temporary pause of study and potential delay of future site activations (for Phase III) may be inevitable. However, inheriting the advantages of the adaptive 2-in-1 designs, the SDDO framework is much more efficient than conventional designs that conduct sequential Phase II and Phase III trials, in terms of \textit{overall} trial duration and sample size. In addition, unblinding data for adaptation decision-making and the concurrent enrollment of ``optimized" treatment arm and control arm may require independent data monitoring committee or independent statistical center. Lastly, robust infrastructure, sufficient resource allocation, and efficient information circulation are needed to avoid logistical-related problems for the supply of experimental agents and to ensure quality and scientific integrity of the study, as well as validity of the interpretation of final results \citep{cerqueira2020adaptive}. Statisticians are expected to take a leading role in such discussions with project teams and health regulators at the trial planning stage, e.g. via a Type-B meeting, details of optimization strategy and Phase II/III trial design should be aligned with health regulators, to further reduce the operational hurdle\citep{dejardin2024dose}.

%\sout{There is an inevitable pause in enrollment and delay in opening the number of sites required for Phase III until the adaptation decision is made;}

The risk of inflating overall Type I error from the adaptation decision of advancing the ``promising'' dose level at interim analysis can be well controlled through careful choices of the tuning parameters $\tau_0$, $\tau_1$, $s_0$, and $s_1$ (Appendix Section \ref{sec:type1}). The adaptive decision-making relies on the joint distribution of surrogate endpoint ORR and final primary endpoint OS with priors associated with dose levels. Our proposed priors can inclusively incorporate the preliminary efficacy information from early-phase studies -- even if such efficacy information is limited, it is still clinically relevant to borrow information from early-phase to increase the overall success rate of late-phase studies (Appendix Section \ref{add_sim}). Extensions of adaptation decisions based on different surrogate and primary endpoints, and priors that incorporate both safety and preliminary efficacy information, are worth further explorations.

% discussion of practicability
The SDDO framework includes multiple enhancements of the original 2-in-1 adaptive design which has been implemented and considered in oncology studies over the years: the first published example is the INDUCE-3 study for head and neck squamous cell carcinoma (Hensen et al., 2020\citep{hansen2020induce}); in Li et al. (2022)\citep{Li2022seamless2in1}, the authors explored the option of a flexible 2-in-1 design for a relapsed/refractory multiple myeloma study; in Cong and Rubin (2023)\citep{chen2023adaptive}, the authors mentioned a possible extension from 2-in-1 to a 3-in-1 design using the SKYSCRAPER-06 study for first-line non-small cell lung cancer as an example. All the aforementioned examples indicate that it is totally feasible to consider such novel adaptive designs in real-world oncology trials. Compared with conventional designs, we acknowledge that the proposed SDDO framework is relatively complicated; however, we strongly believe in its practical use as it is an efficient design option to address key questions outlined in the clinical regulatory projects recently initiated by the FDA Oncology Center of Excellence such as Project Optimus and Project FrontRunner. Whenever needed, sponsors who plan to implement this novel approach can utilize the Agency’s “Complex Innovative Trial Designs (CIDs)” program for consultation (which includes “evaluating Bayesian approaches for the potential to increase clinical trial efficiency”).

\bibliography{seamless.bib}
\bibliographystyle{unsrtnat} 

\newpage

\newpage

\appendix

\section{Tuning Parameter Selection}

\begin{figure}[ht]
    \centering 
     \includegraphics[width=0.6\textwidth]{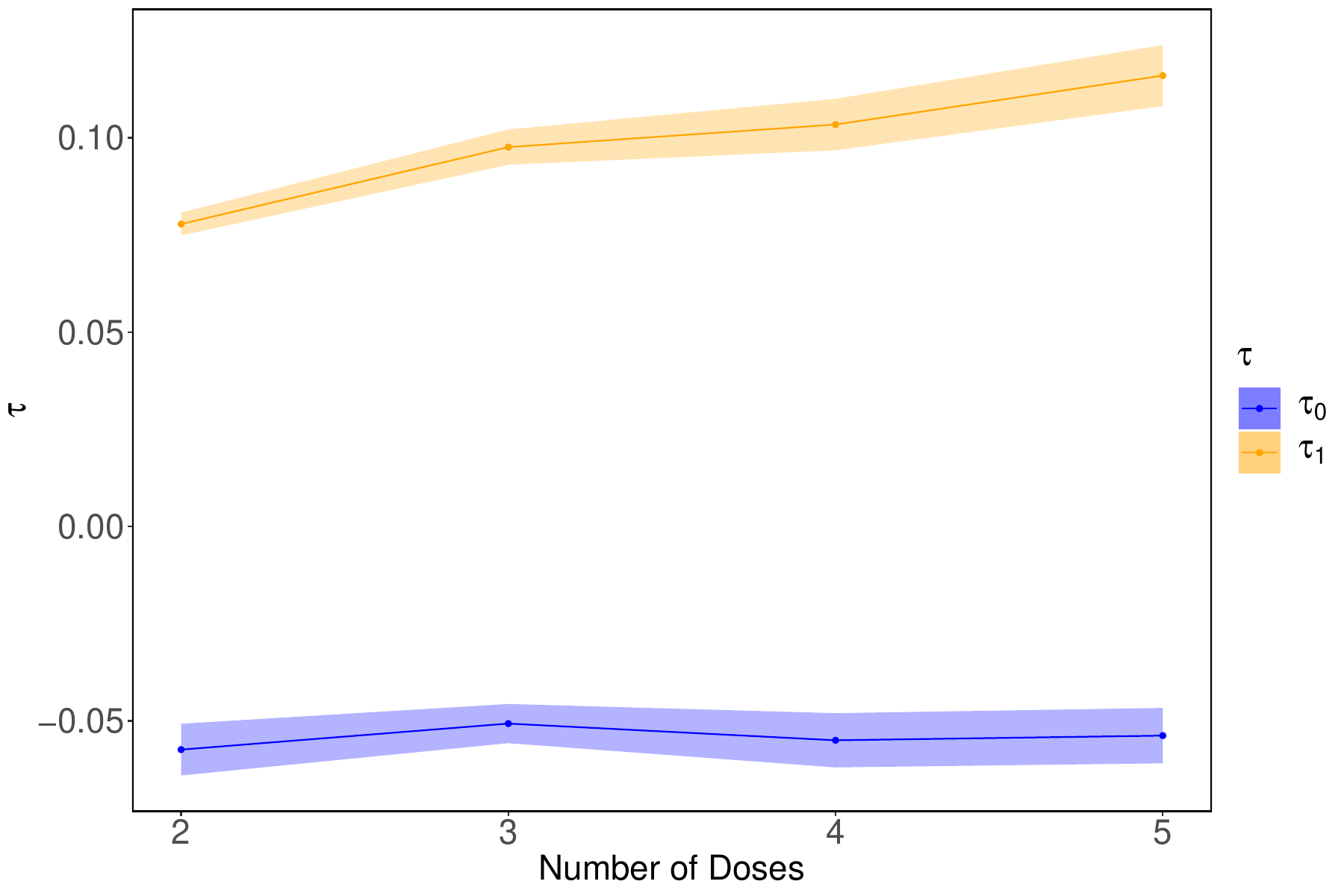}
%\label{flowchart}
\caption{Mean and standard deviation of $\tau_0$ and $\tau_1$ with respect to different numbers of dosages based on 1000 simulations. $\tau_0$ is selected to control the error rate of false negative decision at 5\% with an ORR difference of 0.15. $\tau_1$ is selected to control the error rate of false positive decision at 2.5\%. The prior distribution is set to be uniform for all dosages.}
\label{tau}
\vspace{-3mm}
\end{figure}

Figure \ref{tau} shows the selected value of $\tau_0$ and $\tau_1$ from grid search, using the procedure described in Section \ref{sec:decision}. Notably, $\tau_0$ tends to remain consistent across varying numbers of candidate dosages. This stability can be attributed to the ORR difference of 0.15 being significant enough to distinguish the optimal dose arm from the control arm, and the influence of additional candidate dosages on the value of $\tau_0$ is minimal. In contrast, the value of $\tau_1$ would increase as the number of dosages increases. This finding implies the necessity to apply a more stringent multiple testing correction procedure as the number of dosages increases, for the purpose of controlling false positives.  
 
\section{Type I Error Control}
\label{sec:type1}
We prove that the proposed SDDO design could control the overall Type I error in this section. Without the loss of generality, suppose the desired Type I error is 0.025. To avoid the influence of prior distribution, we assume uniform prior for both ORR and $\text{HR}_\text{OS}$. 

Under the null hypothesis, suppose the probability of terminating the trials, entering Phase II and Phase III study after interim analysis are $a_1,a_2,a_3$ respectively ($a_1+a_2+a_3=1$), and the false positive rate after entering Phase II and Phase III study are denoted as $\alpha_2,\alpha_3$. Therefore, the overall Type I error $\alpha$ could be computed as
$\alpha=a_2\alpha_2+a_3\alpha_3.$
Notice that $\tau_1$ is selected to control the error rate of entering Phase III to be 2.5\% under the null hypothesis, thus we could specify $a_3=0.025$. Additionally, since Phase II study is performed following the pre-planned event size and is not influenced by interim analysis, the Type I error would be controlled at the targeted level, thus $\alpha_2=0.025$. Combining the above result, we have
\vspace{-3mm}
\begin{align*}
\alpha & =a_2\alpha_2+a_3\alpha_3 \\
       & = 0.025a_2+0.025\alpha_3.
\end{align*}

\begin{figure}[h]
    \centering 
     \includegraphics[width=0.5\textwidth]{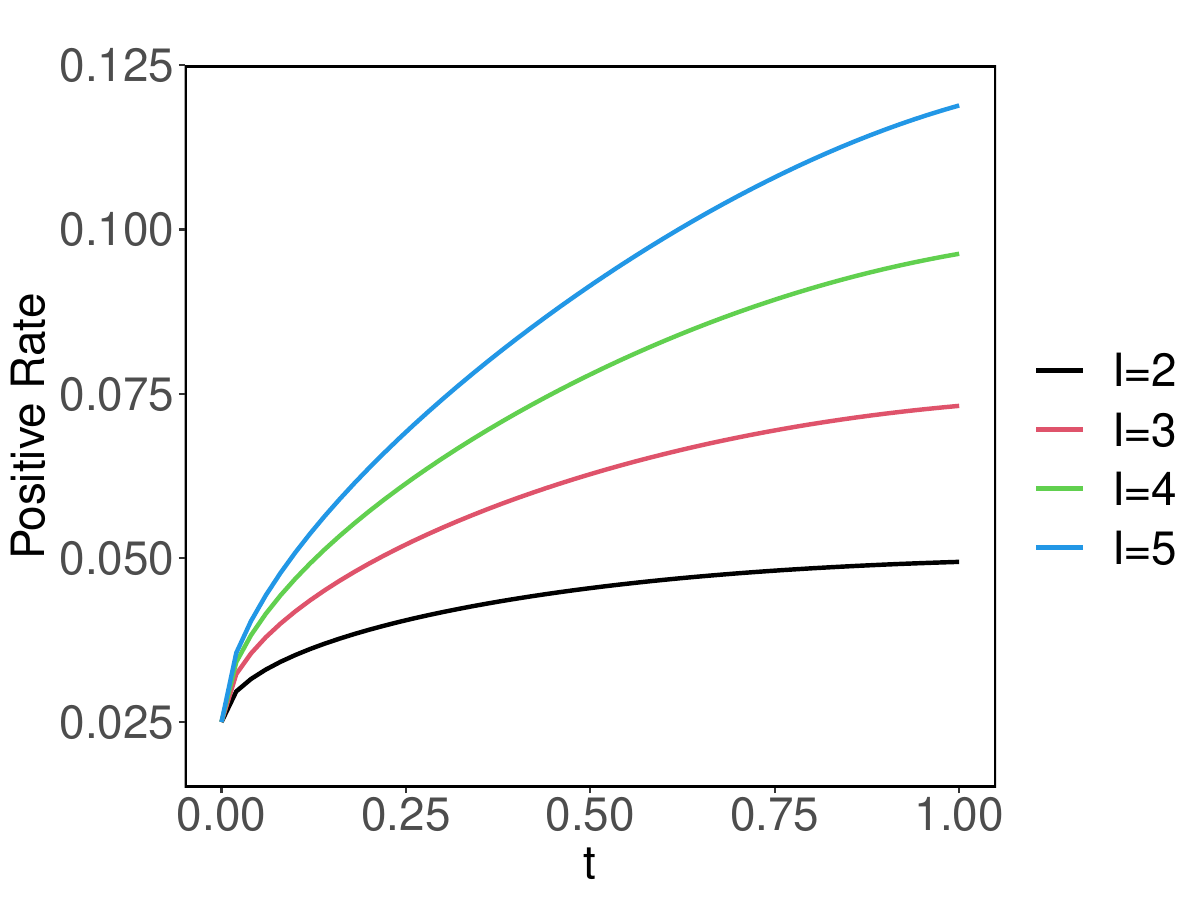}
%\label{flowchart}
\caption{Upper bound of $\alpha_3$ as affected by varying the number of candidate dosages and different values of $t$.}
\label{alpha3_fig}
\end{figure}

Therefore, we proceed to find the upper bound for $a_2$ and $\alpha_3$. We first focus on the upper bound for $\alpha_3$. Based on the SDDO framework, suppose we have $I$ candidate dosage at the beginning of Phase II study. we denote $t$ as the proportion of event size available at interim analysis, $Z(t), Z(1-t)$ refers to the standardized test statistics of the selected optimal dose based on data before and after interim analysis, suppose the clinical trials entering Phase III under null hypothesis we have
\begin{align}
\begin{split}
\alpha_3 & = P(Z(1)\leq Z_{\alpha})\\
 & =P\Big(\sqrt{t} Z(t)+\sqrt{1-t}Z(1-t)\leq Z_{\alpha}\Big) \\
 & \leq P\Big(\sqrt{t} \min (Z_1,Z_2,...,Z_I)+\sqrt{1-t}Z(1-t)\leq Z_{\alpha}\Big) \quad   (Z_1,Z_2,...Z_I \stackrel {i.i.d}\sim N(0,1), Z(1-t)\sim N(0,1) )\\
 & =P\Big(\min (Z_1,Z_2,...,Z_I)\leq \frac{Z_{\alpha}-\sqrt{1-t}Z(1-t)}{\sqrt{t}}\Big)\\
 & =\int ^{\infty}_{-\infty}P\Big(\min (Z_1,Z_2,...,Z_I)\leq \frac{Z_{\alpha}-\sqrt{1-t}z_0}{\sqrt{t}} \Big| Z(1-t)=z_0\Big)f_{Z(1-t)}(z_0)dz_0\\
 & = \int ^{\infty}_{-\infty} \Big[1-\Big(1-\Phi \big(\frac{Z_{\alpha}-\sqrt{1-t}z_0}{\sqrt{t}}\big)\Big)^I \Big]\phi(z_0)dz_0,
 \end{split}
 \label{alpha3}
\end{align}
where $\Phi,\phi$ denote the cdf and pdf of standard normal distribution. Note that Equation \ref{alpha3} illustrates how the upper bound of $\alpha_3$ is influenced by both $t$ and the number of candidate dosages, which need to be determined prior to the start of the clinical study. Therefore, $\alpha_3$ can be more effectively controlled by appropriately specifying values for $t$ and $I$. Figure \ref{alpha3_fig} demonstrates such relationship, clearly showing that 
\begin{equation}
\alpha_3\leq 0.125
\label{alpha3_bd}
\end{equation}
across all examined scenarios.

We then proceed to bound $a_2$. Recall that $a_2$ refers to the probability of entering Phase II study after interim analysis, thus $a_2=P\big(P(p_{i^*}-p_0 \geq \tau_0)>0.9\big)-a_3=P\big(P(p_{i^*}-p_0 \geq \tau_0)>0.9\big)-0.025$. Notice that under null hypothesis $\theta_{i^*}=0$, thus it wouldn't influence the probability $P(p_{i^*}-p_0 \geq \tau_0)$, we omit the conditional part in Equation \eqref{futility} for simplicity.

Under the null hypothesis, at interim analysis of the proposed SDDO framework, we have posterior distribution $p_0\sim \mathcal{N}(\hat \mu_0,\sigma^2)$,  $p_{i^*}\sim \mathcal{N}(\hat \mu_{i^*},\sigma^2)$, where $\sigma=\sqrt{\frac{\hat \mu_0(1-\hat \mu_0)}{n_0}}$, $\hat \mu_0=\frac{y_0}{n_0}, \hat \mu_{i^*}=\frac{y_{i^*}}{n_0}$, $n_0$ is the number of participants enrolled in each dose arm and control arm, and $y_i$ is the number of responders in each dose arm. Therefore, $p_{i^*}-p_0 \sim \mathcal{N}(\hat \mu_{i^*}-\hat \mu_0,2\sigma^2)$, we then have 
\begin{align*}
P\big(P(p_{i^*}-p_0 \geq \tau_0)>0.9\big) & = P\Big(1-\Phi\Big(\frac{\tau_0- (\hat \mu_{i^*}- \hat \mu_0)}{\sqrt{2\sigma^2}}\Big) >0.9\Big) \\
& = P\Big( \Phi\Big(\frac{\tau_0- (\hat \mu_{i^*}- \hat \mu_0)}{\sqrt{2\sigma^2}}\Big)<0.1\Big) \\
& = P\Big( \frac{\tau_0- (\hat \mu_{i^*}- \hat \mu_0)}{\sqrt{2\sigma^2}} <-1.28\Big) \quad (\text{$\Phi^{-1}(0.1)=-1.28$}) \\
& = P\Big(\tau_0- (\hat \mu_{i^*}- \hat \mu_0) < -1.28\sqrt{2\sigma^2}\Big) \\
& =  P\Big( (\hat \mu_{i^*}- \hat \mu_0) > \tau_0+1.28\sqrt{2\sigma^2}\Big).
\end{align*}

Notice that $\hat \mu_{i^*}=\max(\hat \mu_1,\hat \mu_2,...,\hat \mu_I)$ and $\hat \mu_0,\hat \mu_1,\hat \mu_2,...,\hat \mu_I \stackrel {i.i.d}\sim N(p_0,\sigma^2)$, therefore, 
\begin{align*}
\begin{split}
P\big(P(p_{i^*}-p_0 \geq \tau_0)>0.9\big) & =  P\Big( (\hat \mu_{i^*}- \hat \mu_0) > \tau_0+1.28\sqrt{2\sigma^2}\Big)\\
& = 1 - P\Big( (\hat \mu_{i^*}- \hat \mu_0) \leq \tau_0+1.28\sqrt{2\sigma^2}\Big)\\
& = 1-\int^{\infty}_{-\infty} P\Big(\hat \mu_{i^*}\leq \mu+\tau_0+ 1.28\sqrt{2\sigma^2} \Big|\hat \mu_0=\mu  \Big)f_{\hat \mu_0}(\mu) d\mu \\
& = 1-\int^{\infty}_{-\infty}\Phi^I\Big(\frac{\mu+\tau_0+1.28\sqrt{2\sigma^2}}{\sigma}\Big)\phi\Big(\frac{\mu-p_0}{\sigma} \Big)d\mu.
\end{split}
\label{a2}
\end{align*}
Therefore, 
\begin{align}
\begin{split}
a_2 & = P\big(P(p_{i^*}-p_0 \geq \tau_0)>0.9\big)-0.025 \\
    & = 0.975 - \int^{\infty}_{-\infty}\Phi^I\Big(\frac{\mu+\tau_0+1.28\sqrt{2\sigma^2}}{\sigma}\Big)\phi\Big(\frac{\mu-p_0}{\sigma} \Big)d\mu.  
\end{split}
\end{align}

\begin{figure}[h]
    \centering 
     \includegraphics[width=0.9\textwidth]{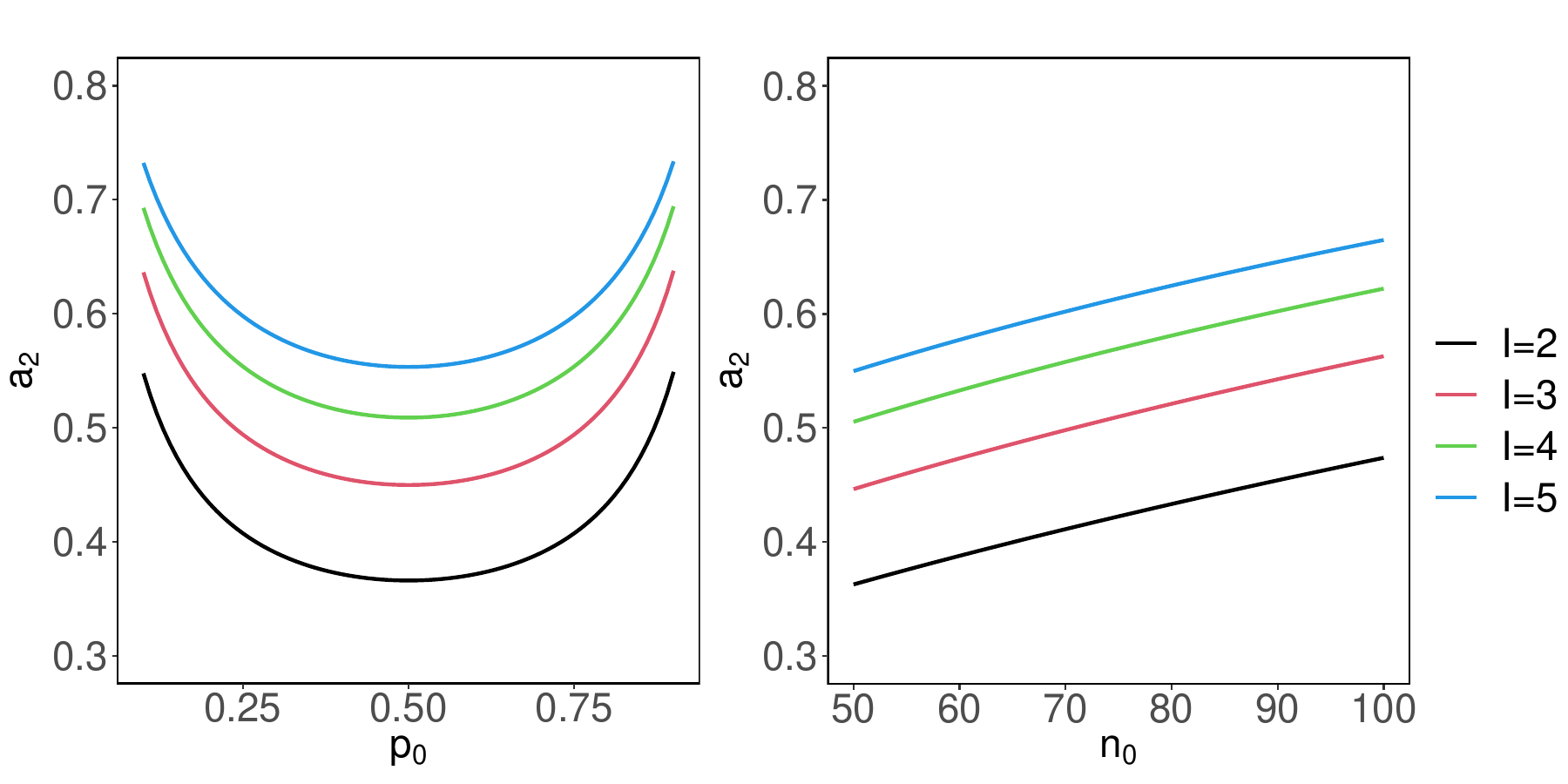}
%\label{flowchart}
\caption{Value of $a_2$ as affected by varying the number of candidate dosages and different values of $p_0$ and $n_0$. The left panel assumes $n_0=80$, and the right panel assumes $p_0=0.2$.}
\label{a2_fig}
\end{figure}

From Equation \ref{a2}, it is evident that the value of \(a_2\) is influenced by three factors: the number of candidate dosages \(I\), the sample size prior to interim analysis \(n_0\), and the ORR under the null hypothesis \(p_0\). Figure \ref{a2_fig} illustrates their relationship, which shows that 
\begin{equation}
   a_2<0.8 
\label{a2_bd}
\end{equation}
across all scenarios. Combining \eqref{alpha3_bd} and \eqref{a2_bd}, the overall Type I error is controlled by
\begin{align*}
\alpha &  = a_2\times 0.025+0.025\alpha_3 \\
  & < 0.025(0.125+0.8) <0.025.
\end{align*}
Therefore, the overall Type I error is controlled for the proposed SDDO design.

\section{Additional Simulation Results} 
\label{add_sim}

\begin{figure}[ht]
    \centering 
     \includegraphics[width=0.95\textwidth]{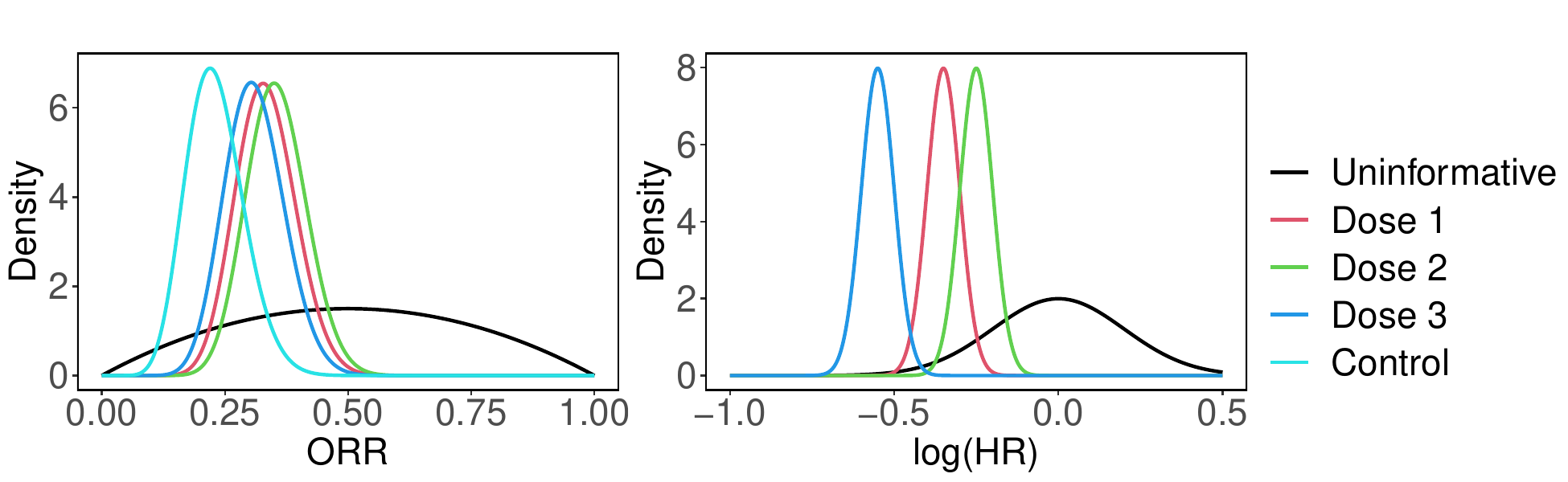}
%\label{flowchart}
\caption{Prior distributions for each dose level on ORR and log hazard ratio corresponding to the simulation scenario in Table \ref{prior_simu}. The uninformative prior is designated as \( \text{Beta}(2,2) \) for ORR, and \( \mathcal{N}(0,0.5) \) for \( \log(\text{HR}_{\text{OS}}) \) across all dosages. The informative prior for ORR follows the distribution \( \text{Beta}(\alpha(d),20) \), where \( \alpha(d) \) is a quadratic function of the dose level. Similarly, the informative prior for \( \log(\text{HR}) \) is given by \( \mathcal{N}(\mu(d),0.05) \), where \( \mu(d) \) also represents a quadratic function of the dose level.}
\label{prior_den}
\end{figure}

\begin{figure}[h]
    \centering 
     \includegraphics[width=0.95\textwidth]{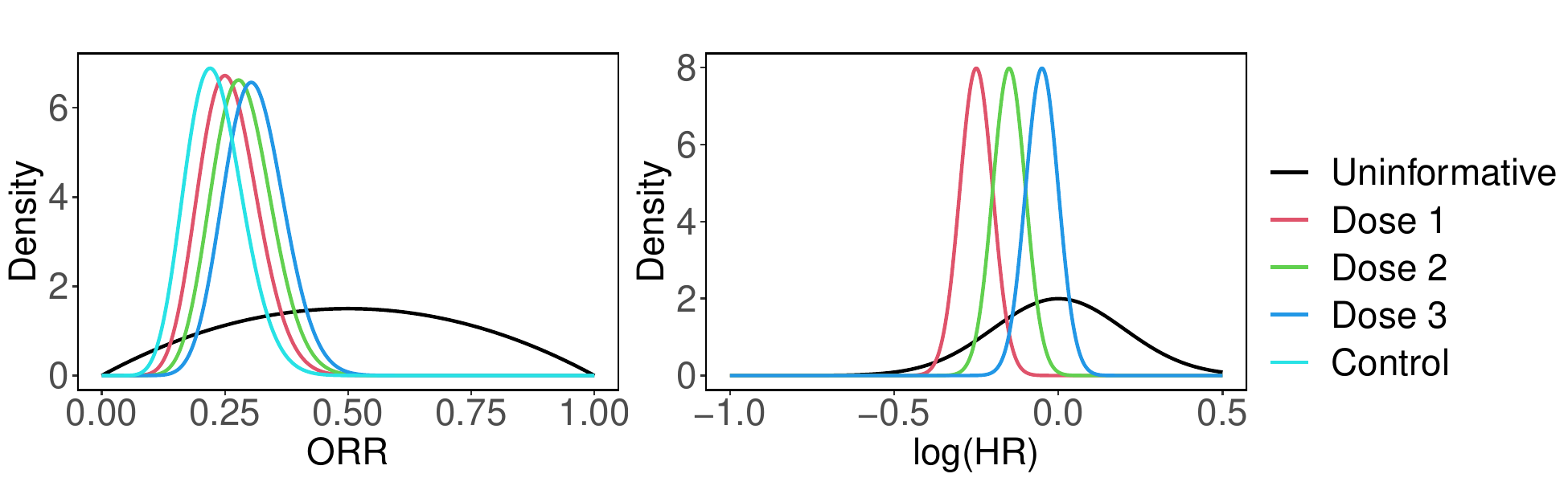}
%\label{flowchart}
\caption{Prior distributions for each dose level on ORR and log hazard ratio corresponding to the simulation scenario in Table \ref{weak_prior_simu}. The uninformative prior is designated as \( \text{Beta}(2,2) \) for ORR, and \( \mathcal{N}(0,0.5) \) for \( \log(\text{HR}_{\text{OS}}) \) across all dosages. The informative prior for ORR follows the distribution \( \text{Beta}(\alpha(d),20) \), where \( \alpha(d) \) is a linear function of the dose level. Similarly, the informative prior for \( \log(\text{HR}) \) is given by \( \mathcal{N}(\mu(d),0.05) \), where \( \mu(d) \) also represents a linear function of the dose level.}
\label{weak_prior}
\end{figure}

In this section, we evaluate the performance of the SDDO framework with respect to various prior distributions assigned to ORR and \( \text{HR}_{\text{OS}} \). Tables~\ref{prior_simu} and~\ref{weak_prior_simu} demonstrate the operating characteristics under both uninformative and informative priors across different scenarios. Specifically, in Table~\ref{prior_simu}, we assume a bell-shaped dose-response relationship, while in Table~\ref{weak_prior_simu}, a monotonic relationship is assumed. For the informative prior, prior distributions that accurately reflect these relationships were assigned to each dose level, while the uninformative prior remains consistent across all dosage levels. Detailed illustrations of the prior distributions are provided in Figures~\ref{prior_den} and~\ref{weak_prior}.

Tables~\ref{prior_simu} and~\ref{weak_prior_simu} clearly demonstrate that the proposed SDDO framework achieves a higher probability of selecting the true optimal dosage when aided by an informative prior derived from data from previous clinical studies, such as Phase I dose-escalation trials. This improvement is attributed to the enhanced decision-making certainty provided by the informative prior, which reinforces the signal of treatment effects through the integration of prior knowledge regarding the outcomes of interest. Additionally, as shown in Table~\ref{prior_simu}, when the treatment effect is significant, the probability of directly advancing to a Phase III study is substantially higher with an informative prior (83.48\% vs 70.43\%), whereas this probability tends to decline when the treatment effect is insignificant as shown in Table~\ref{weak_prior_simu} (13.19\% vs 26.53\%). These observations also suggest that informative priors could significantly improve the probability of making a preferable choice regarding trial expansion, thereby facilitating the ``quick to win, fast to fail'' principle in decision-making.

Besides improving the probability of making optimal decisions regarding dose selection and trial expansion, a smaller event size is required under both scenarios with informative priors while achieving better accuracy at the final analysis. When the treatment effect is significant as in Table~\ref{prior_simu}, the reduction in event size can be attributed to the informative prior strengthening the original signal by accommodating additional information, thus achieving a higher PPoS at the interim analysis for Phase III expansion which could, in turn, contribute to a reduction in the required event size at the final analysis. Meanwhile, when the treatment effect is insignificant as in Table~\ref{weak_prior_simu}, such reduction is mainly due to the lower probability of entering the Phase III study. These observations further indicate that the proposed SDDO design could optimize the cost efficiency of the clinical study significantly.

\begin{table}[ht]
\centering
\footnotesize % Make the font size smaller
\setlength{\tabcolsep}{2pt}
\begin{tabular}{|l|c|c|c|c|c|c|}
\hline
\multicolumn{1}{|c|}{\begin{tabular}[c]{@{}c@{}}Simulation \\ Scenarios\end{tabular}} & \begin{tabular}[c]{@{}c@{}}Optimal Dose\\ Percentage\end{tabular} &  & \begin{tabular}[c]{@{}c@{}}Interim \\ Decision\end{tabular} & \begin{tabular}[c]{@{}c@{}}Positive \\ Rate\end{tabular} & \begin{tabular}[c]{@{}c@{}}Expected \\ Event Size\end{tabular} & \begin{tabular}[c]{@{}c@{}}Expected \\ Study Duration\end{tabular} \\ \hline
\multirow{4}{*}{\begin{tabular}[c]{@{}l@{}}Uninformative Prior:\\ ORR = {[}0.2,0.32,0.35,0.3{]}\\ $\text{HR}_{\text{PFS}}$={[}1,0.7,0.65,0.75{]}\\ $\text{HR}_{\text{OS}}$={[}1,0.7,0.65,0.75{]}\end{tabular}} & \multirow{4}{*}{{[}26.80\%,58.63\%,14.57\%{]}} & Overall & -- & 86.17\% (52.17\%) & 313 & 40.01\\ \cline{3-7} 
 &  & Terminate & 0.56\% & 0 & 72 & 18.84 \\ \cline{3-7} 
 &  & Phase II & 29.01\% & 61.43\% (37.38\%) & 140 & 33.72 \\ \cline{3-7} 
 &  & Phase III & 70.43\% & 97.04\% (58.62\%)& 386 & 42.76 \\ \hline
\multirow{4}{*}{\begin{tabular}[c]{@{}l@{}}Informative Prior:\\ ORR = {[}0.2,0.32,0.35,0.3{]}\\ $\text{HR}_{\text{PFS}}$={[}1,0.7,0.65,0.75{]}\\ $\text{HR}_{\text{OS}}$={[}1,0.7,0.65,0.75{]}\end{tabular}} & \multirow{4}{*}{{[}23.11\%,68.98\%,7.91\%{]}} & Overall & -- & 89.49\% (63.78\%)& 253 & 39.28 \\ \cline{3-7} 
 &  & Terminate & 0.01\% & 0 & 75 & 24.60 \\ \cline{3-7} 
 &  & Phase II & 16.51\% & 62.83\% (45.51\%) & 140 & 33.86 \\ \cline{3-7} 
 &  & Phase III & 83.48\% & 94.76\% (67.41\%)& 276 & 40.36 \\ \hline
\end{tabular}
\caption{Operating characteristics of the SDDO framework based on 10,000 simulation runs under uninformative prior and informative prior. Dose response relationship follows a bell curve under this scenario,  detailed illustrations of the prior distributions are provided in Figure \ref{prior_den}. ``Positive Rate" column shows overall positive effect rate, and in brackets, the rate of selecting the optimal dose with a positive effect.}
\label{prior_simu}
\end{table}

\begin{table}[ht]
\centering
\footnotesize % Make the font size smaller
\setlength{\tabcolsep}{2pt}
\begin{tabular}{|l|c|c|c|c|c|c|}
\hline
\multicolumn{1}{|c|}{\begin{tabular}[c]{@{}c@{}}Simulation \\ Scenarios\end{tabular}} & \begin{tabular}[c]{@{}c@{}}Optimal Dose\\ Percentage\end{tabular} &  & \begin{tabular}[c]{@{}c@{}}Interim \\ Decision\end{tabular} & \begin{tabular}[c]{@{}c@{}}Positive \\ Rate\end{tabular} & \begin{tabular}[c]{@{}c@{}}Expected \\ Event Size\end{tabular} & \begin{tabular}[c]{@{}c@{}}Expected \\ Study Duration\end{tabular} \\ \hline
\multirow{4}{*}{\begin{tabular}[c]{@{}l@{}}Uninformative Prior:\\ ORR = {[}0.2,0.25,0.27,0.3{]}\\ $\text{HR}_{\text{PFS}}$={[}1,0.95,0.9,0.85{]}\\ $\text{HR}_{\text{OS}}$={[}1,0.95,0.9,0.85{]}\end{tabular}} & \multirow{4}{*}{{[}14.06\%,29.10\%,56.84\%{]}} & Overall & -- & 19.13\% (14.91\%) & 221 & 31.69\\ \cline{3-7} 
 &  & Terminate & 7.44\% & 0 & 78 & 18.49 \\ \cline{3-7} 
 &  & Phase II & 66.03\% & 11.83\% (8.78\%) & 140 & 29.73 \\ \cline{3-7} 
 &  & Phase III & 26.53\% & 42.67\% (34.31\%)& 462 & 40.26 \\ \hline
\multirow{4}{*}{\begin{tabular}[c]{@{}l@{}}Informative Prior:\\ ORR = {[}0.2,0.25,0.27,0.3{]}\\ $\text{HR}_{\text{PFS}}$={[}1,0.95,0.9,0.85{]}\\ $\text{HR}_{\text{OS}}$={[}1,0.95,0.9,0.85{]}\end{tabular}} & \multirow{4}{*}{{[}9.62\%,25.70\%,64.68\%{]}} & Overall & -- & 18.13\% (14.68\%) & 185 & 30.98 \\ \cline{3-7} 
 &  & Terminate & 2.54\% & 0 & 80 & 20.01 \\ \cline{3-7} 
 &  & Phase II & 84.27\% & 13.13\% (10.36\%) & 140 & 29.86 \\ \cline{3-7} 
 &  & Phase III & 13.19\% & 53.52\% (45.11\%) & 495 & 40.19 \\ \hline
\end{tabular}
\caption{Operating characteristics of the SDDO framework based on 10,000 simulation runs under uninformative prior and informative prior. Treatment effect is assumed to be insignificant for all dosages, detailed illustrations of the prior distributions are provided in Figure \ref{weak_prior}. ``Positive Rate" column shows overall positive effect rate, and in brackets, the rate of selecting the optimal dose with a positive effect.}
\label{weak_prior_simu}
\end{table}

%Table \ref{prior_simu} and \ref{weak_prior_simu} clearly show that, with the aid of an informative prior drawing on data from previous clinical studies such as Phase I dose-escalation trials, the proposed SDDO framework achieves a higher probability of selecting the true optimal dosage (68.98\% vs 58.63\%) and directly advancing to Phase III studies (83.48\% vs 70.43\%). This improvement is attributed to the fact that an informative prior can enhance decision-making certainty by reinforcing the signal of treatment effects through the integration of prior knowledge on the outcomes of interest. Additionally, the expected event size upon entering Phase III studies is significantly reduced when utilizing an informative prior (276 vs 386) while maintaining a comparable level of statistical power (94.76\% vs 97.04\%). This reduction can also be attributed to the informative prior strengthening the original signal by accommodating additional information, thus achieving a higher PPoS at interim analysis which could, in turn, contribute to a reduction in the required sample size at the final analysis.

\end{document}